\documentclass[preprint]{elsarticle}

\usepackage{hyperref}

\usepackage[ margin=1in]{geometry}
\usepackage{enumitem}

\usepackage{amsmath,amsfonts,amsthm,bm}
\usepackage{booktabs}
\usepackage{subcaption}
\usepackage{placeins}
\usepackage{epsfig}
\usepackage{enumitem}
\usepackage{xcolor}
\usepackage{tabularx}
\usepackage{multirow}
\usepackage{natbib}
\newcolumntype{Y}{>{\centering\arraybackslash}X}
\usepackage{soul}

\usepackage{tikz}
\usetikzlibrary{shapes,arrows,fit,positioning,calc}
\tikzstyle{io2} = [rectangle, minimum height=4em, minimum width=13em, text centered, draw=black, fill=yellow!3]
\tikzstyle{model} = [rectangle, draw, fill=red!3, 
    text width=25em, text centered, rounded corners, minimum height=4em, minimum width = 27.5em]
    \tikzstyle{model2} = [rectangle, draw, fill=red!3, 
        text width=11em, text centered, rounded corners, minimum height=4em, minimum width = 13em]
\tikzstyle{arrow} = [thick,->,>=stealth]

\journal{Journal of \LaTeX\ Templates}

\begin{document}

\begin{frontmatter}

\title{A Statistical Simulation Method for Joint Time Series of Non-stationary Hourly Wave Parameters}

\author[tud]{W.S.~J\"ager\corref{cor1}}
\ead{w.s.jager@tudelft.nl}
\author[tum]{T. Nagler}
\author[tum]{C. Czado}
\author[deltares]{R.T. McCall}

\cortext[cor1]{Corresponding author}

\address[tud]{Department of Hydraulic Engineering, Delft University of Technology, Delft, The Netherlands}
\address[tum]{Department of Mathematics, Technical University of Munich, Garching, Germany}
\address[deltares]{Department of Marine and Coastal Information Systems, Deltares, Delft, The Netherlands}

\begin{abstract}
Statistically simulated time series of wave parameters are required for many coastal and offshore engineering applications, often at the resolution of approximately one hour. Various studies have relied on autoregressive moving-average (ARMA) processes to simulate synthetic series of wave parameters in a Monte Carlo sense. However, accurately representing inter-series dependencies has remained a challenge. In particular, the relationship between wave height and period statistics is complex, due to the limiting steepness condition. Here, we present a new simulation method for joint time series of significant wave height, mean zero-crossing periods and a directional regime variable. The latter distinguishes between northern and southwestern waves. The method rests on several model components which include renewal processes, Fourier series with random coefficients, ARMA processes, copulas and regime-switching. A particular feature is a data-driven estimate for a wave height-dependent limiting wave steepness condition which is used to facilitate copula-based dependence modeling. The method was developed for and applied to a data set in the Southern North Sea. For this site, the method could simulate time series with realistic annual cycles and inter-annual variability. In the time series data, the bivariate distribution of significant wave height and mean zero-crossing period was well represented. An influence of the directional regime on the bivariate distribution could also be modeled. However, the influence was not as strong in simulated data as in observed data. Finally, simulated series captured duration and inter-arrival time of storm events well. Potential applications for output of the simulation method range from the assessment of coastal risks or design of coastal structures to the  planning and budgeting of offshore operations.
\end{abstract}

\begin{keyword}
Significant wave height, Mean zero-crossing period, ARMA, Copula
%\texttt{elsarticle.cls}\sep \LaTeX\sep Elsevier \sep template
%\MSC[2010] 00-01\sep  99-00
\end{keyword}

\end{frontmatter}

\section{Introduction}

% 1 more paragraph relevance of coastal risk analysis

Many engineering applications call for the generation of synthetic time series of wave conditions, e.g., the simulation of as yet unobserved and possibly unanticipated, high-impact storms \citep[e.g.,][]{VanDongeren2017,Sebastian2017}; the evaluation of long-term morphodynamic impacts of coastal interventions \citep[e.g.,][]{Vitousek2017}; and the planning and safe
execution of offshore operations, where the prediction of calm periods is important \citep[e.g.,][]{leontaris2016probabilistic}.

In principle, sea storms are segments of multivariate temporal processes of metocean variables that pose a hazard to the environment or operation of interest. Typically, these processes are described by hourly statistics, for example, the significant wave height, which is computed from a spectrum of individual waves. The processes exhibit strong state-to-state autocorrelation on short time scales, seasonal cycles on annual and multi-annual time scales, inter-series dependences and, potentially, long-term trends \citep[e.g.,][]{Mendez2006,Mendez2007,Solari2011}. These statistical features make it challenging to model time series of metocean variables, including sea storms. 

Many simulation methods are based on renewal processes to model alternating sequences of storm and calm durations \citep{michele2007multivariate,callaghan2008statistical,corbella2013simulating,li2014probabilisticestimation,wahl2016probabilistic,davies2017Improved}. For the storm periods, high temporal resolution time series of the relevant metocean variables are then derived from an idealized `storm shape'. For most applications hourly values are needed. A typical assumption is that each univariate time series segment corresponds to two sides of a symmetric triangle whose height determines the peak value and whose base is defined by the storm duration. Similar schematizations with alternative geometrical shapes have also been suggested \citep{Martin-Hidalgo2014,Soldevilla2015}. Furthermore, the peak values of different processes are modeled as interdependent, for example using copulas. 

An advantage of this approach is that the modeling effort can be reduced, because features of the metocean time series that are less relevant for the application are not resolved. An example are serial dependence or dependencies between the variables during calm periods. On the other hand, resulting models are application specific, because they rely on predefined storm shapes and critical threshold values,  which are likely to differ per application. For instance, an operating vessel can be sensitive to metocean conditions that a sandy beach is not. An alternative to the methods based on renewal processes is to model complete time series. This increases the modeling burden, but allows for more flexibility in terms of potential applications. 

Currently, three lines of research concentrate on simulating multivariate time series of metocean variables with high temporal resolution. Guanche et al. \cite[][and references therein]{Guanche2013} developed a simulation method based on statistical downscaling. The authors statistically simulate time series of larger-scale sea level pressure fields with autoregressive moving average (ARMA) models from which they then derive local sea state time series. 

Furthermore, ARMA models have been used to directly represent time series of metocean variables at a specific location, most of them at three-hourly scales. Multiple studies exist on univariate time series of significant wave height \citep{Athanassoulis1995,GuedesSoares1996a,GuedesSoares1996,Scotto2000,Stefanakos2006}. Extensions to bivariate processes have been made by including the mean wave periods \citep{GuedesSoares2000} and by including surges \citep{Cai2008}. In addition to significant wave height and peak period, Solari and van Gelder \cite{Solari2011} incorporated parameters related to wind speed, wind direction and wave direction, thus simulating five interrelated processes. The bivariate and multivariate approaches used so-called vector ARMA (VARMA) models, which are able to capture linear interdependencies between multiple time series. However, Solari and van Gelder reported that dependencies could not always be adequately represented. 

Finally, copulas and vine-copulas have been adopted to model both serial dependence as well as inter-series dependencies of metocean processes. For instance, Leontaris et al. \cite{leontaris2016probabilistic} simulated wind speeds and significant wave height, while J\"ager and Morales N\'apoles \cite{jager2017joint} simulated significant wave height and mean zero-crossing periods. A comparative study of a copula-based serial dependence model to an ARMA model for significant wave height time series has been conducted by Solari and van Gelder ~\cite{Solari2011a}. They found that storm frequency and persistence of storms were better represented by the copula-based model, whereas longterm autocorrelation was better represented by the ARMA model. 

%However,  \citep{Jaeger2017} report that persistency characteristics of the simulated time series may not be accurate. 
%However, such time series have been applied to offshore problems, but not yet to coastal applications. They mostly focus on developing the simulation method. 

The above studies used different techniques to account for non-stationarities. The simplest approach has been to focus on the most important season \cite{jager2017joint} or to piecewise model seasons or months \cite{li2014probabilistic,wahl2016probabilistic,leontaris2016probabilistic}. Other studies have used a superposition of linear or cyclic functions of time \cite{Athanassoulis1995,GuedesSoares1996,Stefanakos2006,callaghan2008statistical,Solari2011,corbella2012predicting} and climate indices as co-variates \cite{Mendez2006, Mendez2007,serafin2014simulating,davies2017Improved} to represent trends or seasonal cycles on semiannual to decadal time scales. Climate indices under consideration were the North Atlantic Oscillation (NAO), the Southern Oscillation Index (SOI), the Pacific-North America (PNA) and the El Ni\~no-Southern Oscillation (ENSO) index. Another difference between the techniques is that some studies decompose the metocean processes into seasonal mean process, a seasonal standard deviation process and a stationary process \citep{Athanassoulis1995,GuedesSoares1996,Stefanakos2006,wahl2016probabilistic}, while the others apply non-stationary probability distributions (i.e., distributions with time-varying parameters) \citep{Mendez2006,Mendez2007,callaghan2008statistical,Solari2011,corbella2012predicting,serafin2014simulating,li2014probabilistic,leontaris2016probabilistic,davies2017Improved}.

In this article, we develop a new probabilistic simulation method for joint time series of non-stationary  wave parameters with an hourly resolution. More precisely, the wave parameters are the spectral significant wave height, $H_{m0}$, the mean-zero crossing period, $T_{m02}$, and a directional regime $\Theta$. In the remainder of the article we use this notation for the variables. An overview is given in Table~\ref{tab:nomenclature}.
\begin{table}[ht]
	\caption{Overview of Variables}
	\label{tab:nomenclature}
	\centering
	\begin{tabular}{l c c c}
		\toprule
		Variable & Unit & Name & Sample space \\
		\midrule
		Significant wave height & m & $H_{m0}$ & $\lbrace \mathbb{R}^+ \rbrace$ \\ 
		Mean zero-crossing period & s & $T_{m02}$ & $ \lbrace \mathbb{R}^+ \rbrace$  \\ 
		Wave direction regime & $-$ &  $\Theta$ &  $\lbrace 0,1 \rbrace$ \\ 
		\bottomrule
	\end{tabular}
\end{table}

While we reapply many techniques that have been suggested in the literature cited above, the method presented here distinguishes itself from others
 on five principle points: \begin{enumerate}
 
\item We develop a data-driven equation for the limiting wave steepness condition at the study location and use it for an initial variable transformation. In this way, we separate in our modeling the deterministic part of the relationship between $H_{m0}$ and $T_{m02}$, which is steepness-induced wave breaking, from the stochastic part, which is due to common meteorological and geographical factors.

\item Modeling of the mean wave direction is simplified by assuming a categorical variable with two possible values, north and southwest, which we refer to as wave direction regime. The assumption is reasonable given the geographical context of the measurement station. The main advantage is that we circumvent challenges related to modeling a circular variable and avoid inaccuracies that could arise from ignoring the circular aspect. Time series of the wave direction regime are modeled as a seasonal alternating renewal process, inspired by \cite{Salvadori2006}.

\item Instead of applying a VARMA model with joint-normally distributed residuals to the bivariate time series of $H_{m0}$ and $T_{m02}$, we estimate two univariate ARMA models with a non-normal joint residual distribution constructed via a copula. Recent examples of such an approach for other types of environmental time series can be found in \cite{Erhardt2017,Bevacqua2017}.

\item The wave direction regime is used to trigger regime switches in the joint residual distribution to account for possible differences in the statistical characteristics of northern and south-western waves. Differences are expected, because south-western waves are mostly wind-sea due to a limited fetch length, while northern waves can be a mixture of swells and wind-seas. 

\item Similar to existing studies, we use Fourier series to characterize a seasonal mean process and a standard deviation process. However, we assume that the Fourier coefficients are random variables, potentially dependent, instead of constants in order to represent inter-year differences and dependencies between the processes on yearly time scales. 
\end{enumerate}

The paper is organized as follows. Section~\ref{sec:data} introduces a data set from the measuring station Europlatform in the Dutch Southern North Sea. This data set will be used to develop and illustrate the simulation method. The section also shows how the wave direction affects the bivariate distribution of $H_{m0}$ and $T_{m02}$ and motivates why wave directions are clustered into two regimes. Sections~\ref{sec:model-development} and~\ref{sec:sim_results} represent the core of this article. Section~\ref{sec:model-development} develops the methodology for jointly modeling time series of the wave parameters. Section~\ref{sec:sim_results} shows simulation results. Finally, Section~\ref{sec:discussion} discusses the main limitations of the methods and Section~\ref{sec:conclusion} contains the conclusions. 

\section{Data and Regime Definition} \label{sec:data}

\begin{figure}[thb]
   \centering
   \frame{
   \includegraphics[width=0.4\linewidth]{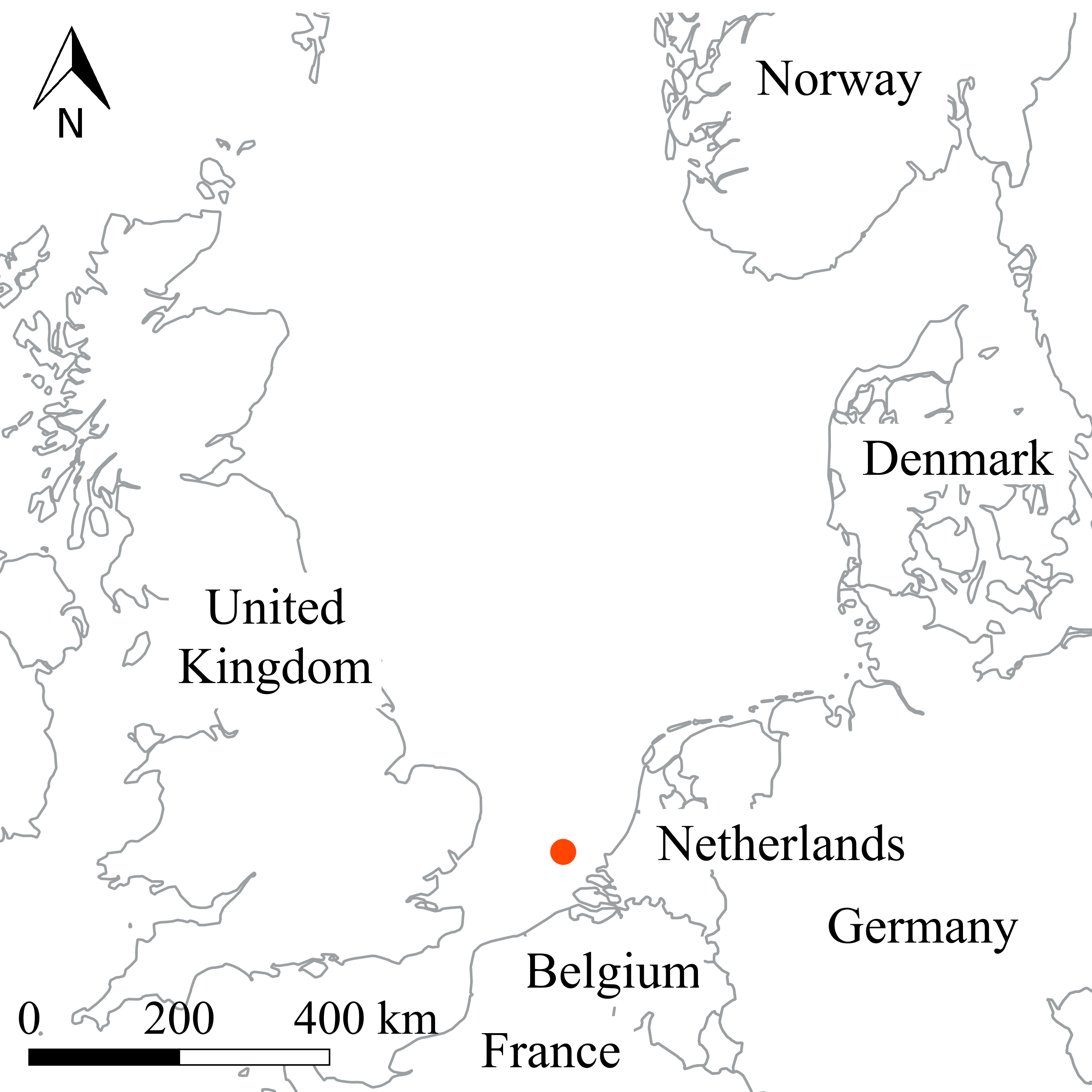}
   }
   \caption{Location of the offshore measuring station ``Europlatform'' indicated as red circle.}
   \label{fig:EPF}
\end{figure}

The data stems from the Europlatform, which is located 38km off the coast of Rotterdam ($52^\circ00'$N, $03^\circ17'E$) at a water depth of approximately $26.$5m (Figure~\ref{fig:EPF}). The measuring station is operated by the Dutch Ministry of Infrastructure and the Environment (in Dutch: Rijkswaterstaat). Our data set consists of hourly measurements of three wave statistics, $H_{m0}$, $T_{m02}$ and the mean wave direction, for a period of 24 years (1 Jan 1991 - 31 Dec 2015). 

The existence of leap years complicates time series modeling on an hourly scale, because the number of hours per year varies. Annual seasonal processes would thus have a period that differs by $24$ hours if the year is a leap year as compared to when it is not. To avoid this, we introduced a new calendar for modeling purposes, in which all years are equally long. This calendar assumes that all Februaries consist of $28$ days and $6$ additional hours. Thus, each year has $8766$ hours, instead of non-leap years having $8760$ hours and leap years having $8784$ hours. Of course, days in different years start at different times of the day, but this is not relevant for our application and has no effect on the results. However, the reader should be aware that dates mentioned and displayed do not exactly correspond to actual dates and times. 

The data coverage is higher than $94\%$; most missing values arise before 2003. The period 1 Jan 2003 - 31 Dec 2014, according to the new calendar, has only five instances in which values of the three wave statistics are jointly missing.  We have filled these by linear interpolation. If a component of the simulation model required a complete time series record without missing values for parameter estimation, this shorter-length record was used  and this is mentioned in the corresponding section. Otherwise, models were estimated based on the full-length record. 

\begin{figure}[thb]
	\centering 
	\begin{subfigure}[t]{0.49\textwidth}
		\includegraphics[width=1\linewidth]{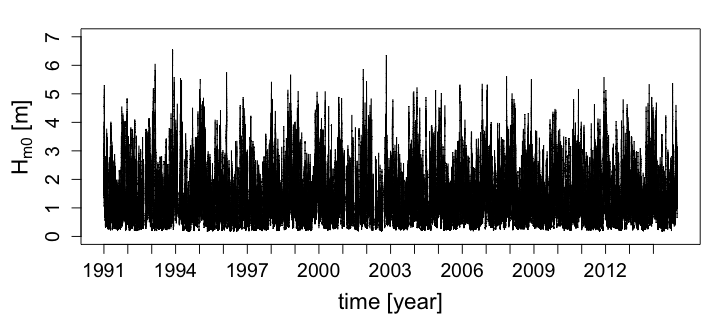}
				\caption{}
		\label{fig:ts_x_hs}
	\end{subfigure} 
	\begin{subfigure}[t]{0.49\textwidth}
		\includegraphics[width=1\linewidth]{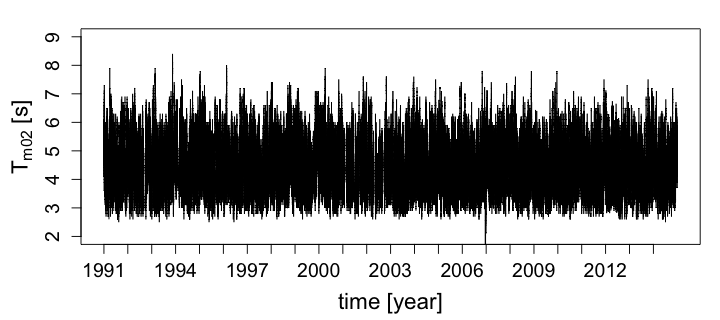}
				\caption{}
		\label{fig:ts_x_tp}
	\end{subfigure}
	\caption{Time series of (a) $H_{m0}$ and (b) $T_{m02}$ from January 2003 to December 2014.  Observations are $1$ hour apart.}
\end{figure}

\begin{figure}[thb]
   \centering
   \includegraphics[width=0.4\linewidth]{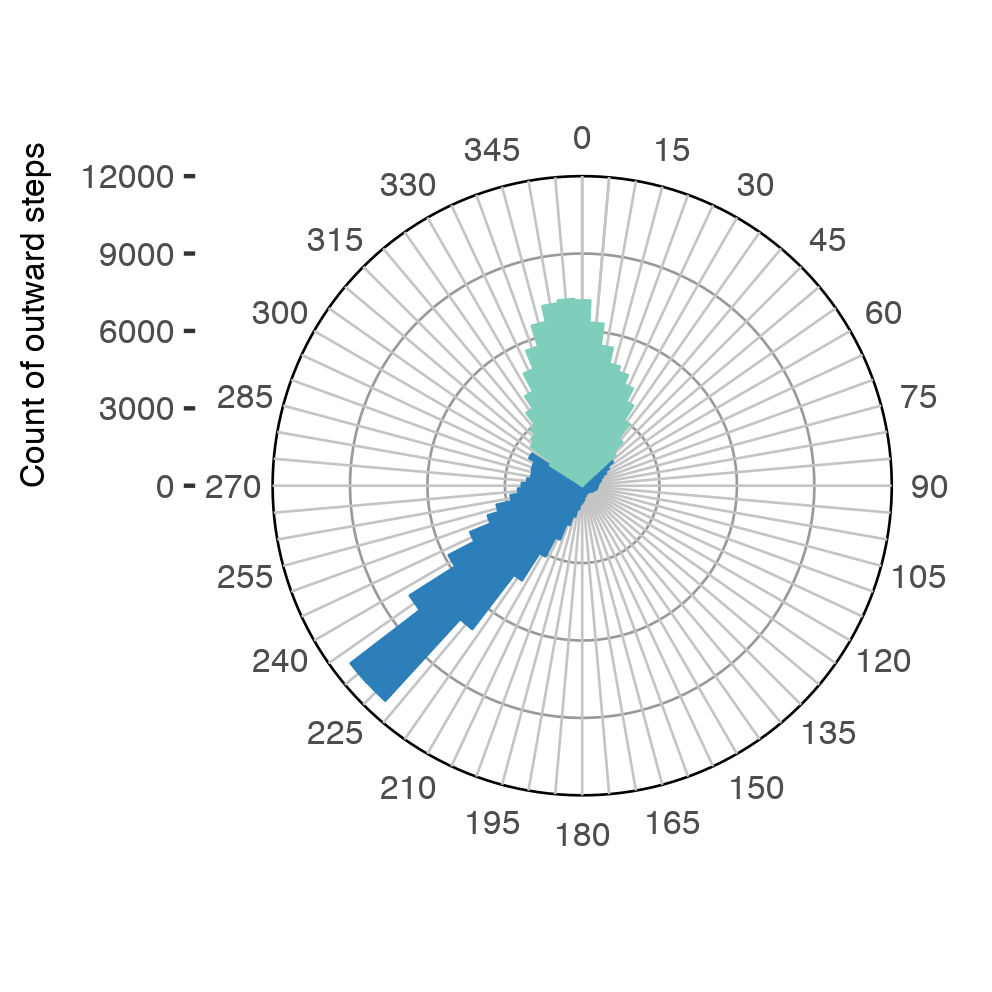}
   \caption{Circular histogram of direction wave directions. Green and blue indicate the two wave direction regimes.}
   \label{fig:histogram-WA}
\end{figure}

The observed time series for $H_{m0}$ and $T_{m02}$ are shown in Figures~\ref{fig:ts_x_hs} and~\ref{fig:ts_x_tp}. Most waves either originate from distinctly northern or south-western directions (Figure~\ref{fig:histogram-WA}). Because of the geographical characteristics of the location (cf. Figure~\ref{fig:EPF}), waves from northern directions can be swells, wind seas or a mixed sea state, while waves from south-western directions are mainly wind seas. For this reason, we expect differences in the statistical properties of northern and south-western waves. 

Issues can arise when ignoring the circular nature of a variable. For a variable in polar coordinates $0^\circ$ and $360^\circ$ are identical directions. Furthermore, $358^\circ$ cannot be called larger than $2^\circ$ and neither is $180^\circ$ their meaningful average. To analyze and model such a variable, many standard statistical tools or measures for ratio variables, such as $H_{m0}$ and $T_{m02}$, might not be suitable. For example, it would be difficult to interpret a rank correlation of the mean wave direction with another variable. If we ignore the circular nature of the mean wave direction and compute its rank correlation with $H_{m0}$, we obtain a value of $0.11$. The calculation assumes that $0^\circ$ is the smallest and $360^\circ$ the largest possible direction. Now suppose that, we would have defined mean wave direction as the direction into which the waves are traveling, contrary to the norm, which is the direction of origin. All our measurements would be shifted by $180^\circ$, but the dependence between mean wave direction and $H_{m0}$ would remain the same. However, we compute a rank correlation value of $-0.21$ for this case. The example shows the danger of obtaining misleading results, when neglecting the circular nature of mean wave direction. 

In the present geographical context it seems natural to represent the wave direction as a categorical variable with two states in order to circumvent issues related to circularity. Hence, the data have been partitioned into two clusters representing the two main directional sectors and a new variable, the wave direction regime, is defined as follows: \begin{equation}
\label{eq:regime}
\Theta_t = \begin{cases}
0, \qquad \mbox{mean wave direction at time \textit{t}} \in (304^\circ, 48^\circ),  \\
1, \qquad \mbox{mean wave direction at time \textit{t}}\in [48^\circ, 304^\circ].
\end{cases}
\end{equation}

\section{Simulation Method}
\label{sec:model-development} 

The statistical simulation method builds on several modeling steps with different probabilistic models. This section motivates and explains the steps rather than describes the underlying probabilistic models. However, the reader can find the main concepts of ARMA processes and copulas explained in~\ref{sec:background-arma} and~\ref{sec:background-copulas}. These appendices also point to introductory literature. 

%, but for more comprehensive introductions the reader is referred to Box et al. \cite{Box2015} and Joe \cite{joe2014dependence}.

\subsection{Overview}

\begin{figure}[thb]
\centering
\resizebox{\textwidth}{!}{
\begin{tikzpicture}[node distance = 1cm]
\node[io2](theta){Mean wave direction};
\node[io2, right = 2cm of theta](Hm0){Significant wave height};
\node[io2,right = 0.5cm of Hm0](Tm02){Mean zero-crossing period};
\node[text = gray] at ($(theta.north west)+(3.5,0.6)$) {\textbf{Input data: Joint time series of wave parameters}};
\draw[gray, thick] ($(theta.north west)+(-1,1)$)  rectangle ($(Tm02.south east)+(1,-0.5)$);
\coordinate (CENTER) at ($(Hm0)!0.5!(Tm02)$);
\node[model,below = 2.75cm of CENTER](steepness){Model for limiting steepness condition (Section~\ref{sec:method-steepness})};
\node[model,below = 4.75cm of CENTER](normalization){Normalization and decomposition into seasonal and stationary components (Section~\ref{sec:method-normalization-deseasonalization})};
\node[model2,below = 6cm of Hm0](stationary){Model for stationary components (Section~\ref{sec:method-stationary})};
\node[model2,below = 6cm of Tm02](seasonal){Model for seasonal components (Section~\ref{sec:method-nonstationary})};
\node[model2,below = 2.05cm of theta](regime-def){Reduction to regime time series (Section~\ref{sec:method_WA})};
\node[model2,below = 6cm of theta](regime-switches){Seasonal model for regime switches (Section~\ref{sec:method_WA})};
\node[text = gray] at ($(regime-def.north west)+(0.5,0.6)$) {\textbf{Modeling steps}};
\draw[gray, thick] ($(regime-def.north west)+(-1,1)$)  rectangle ($(seasonal.south east)+(1,-0.5)$);
\draw [arrow] (theta)  -- (regime-def);
\draw [arrow] (regime-def)  -- (regime-switches);
\draw [arrow] (regime-def)  -- (stationary.north west);
\draw [arrow] (Hm0)  -- (steepness);
\draw [arrow] (Tm02)  -- (steepness);
\draw [arrow] (steepness)  -- (normalization);
\draw [arrow] (normalization)  -- (stationary);
\draw [arrow] (normalization)  -- (seasonal);
\end{tikzpicture}
}
		\caption{Overview of the steps of the statistical simulation method for wave parameter time series}
		\label{fig:modeling_steps}
\end{figure}
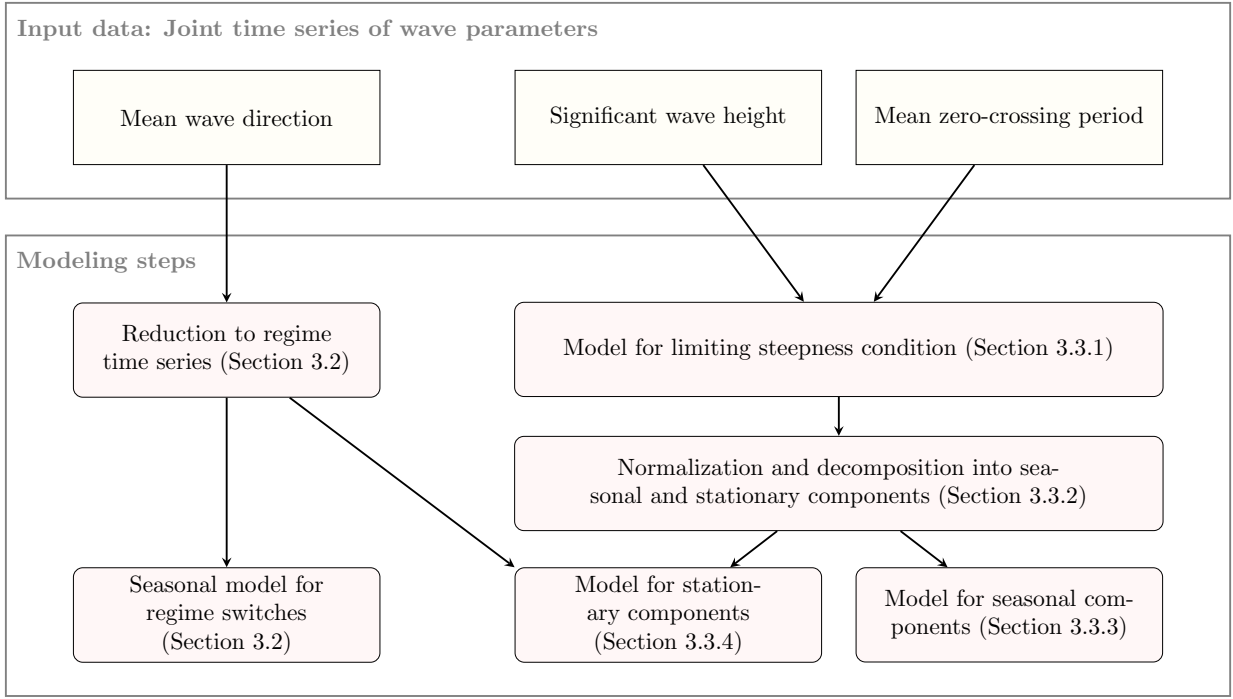

The modeling steps are shown in the flowchart in Figure~\ref{fig:modeling_steps}. First, $\Theta$ is derived according to equation~(\ref{eq:regime}) from the mean wave direction series. Then, the durations for which the directions remain in each regime before switching to the other are represented by a seasonal renewal process (Section~\ref{sec:method_WA}). Independently thereof, a limiting wave steepness condition is estimated for the collected data and used to remove the effects of steepness-induced wave breaking (Section~\ref{sec:method-steepness}). Next, the data of $H_{m0}$ and $T_{m02}$ are normalized and decomposed into stationary and non-stationary processes (Section~\ref{sec:method-normalization-deseasonalization}). The non-stationary processes are modeled using Fourier series with random coefficients (Section~\ref{sec:method-nonstationary}), while the stationary processes are modeled as ARMA using a regime-switching, joint residual distribution (Section~\ref{sec:method-stationary}). The regime switches are triggered by $\Theta$.

\subsection{Wave Angle Regimes}
\label{sec:method_WA}

\begin{figure}[thb]
\centering
		\includegraphics[width=0.4\linewidth]{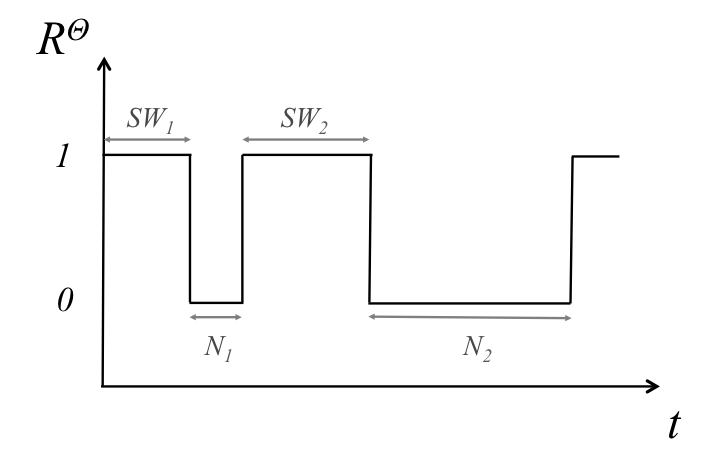}
		\caption{Illustration of renewal process}
		\label{fig:renewal}
\end{figure}

The process $\lbrace \Theta_t \rbrace_t$ is modeled as an alternating binary renewal process, following \cite{Salvadori2006}, who described wet and dry periods of precipitation in this way. For this application, the durations for which waves are coming from one of the two directions are random variables, $N$ and $SW$. For example, the initial direction wave direction regime is $0$ and remains that for a time $SW_1$. Then it switches to $1$ and remains that for a time $N_1$. It is then $0$ for time $SW_2$, and so on.  Figure~\ref{fig:renewal} illustrates the process. 

An alternating renewal process supposes that both sequences $\lbrace N_n \rbrace$ and $\lbrace SW_n \rbrace$, $n \geq 1$, are independent and identically distributed. However, $N_n$ and $SW_n$ may be dependent. We made two modifications to this set up. On one hand, we also allow $SW_n$ to depend on $N_{n-1}$, $n \geq 2$. On the other hand, we assume that both $N_n$ and $SW_n$, $n \geq 1$, depend on the time of the year as well. Thus,  we estimated a bivariate distribution
for $(N_{n-1}, SW_{n})$ and for $(SW_{n},N_{n})$ for each season, using a decomposition into univariate distributions and a copula. 

\begin{figure}[thb]
	\centering 
	\begin{subfigure}[t]{0.75\textwidth}
		\includegraphics[width=1\linewidth]{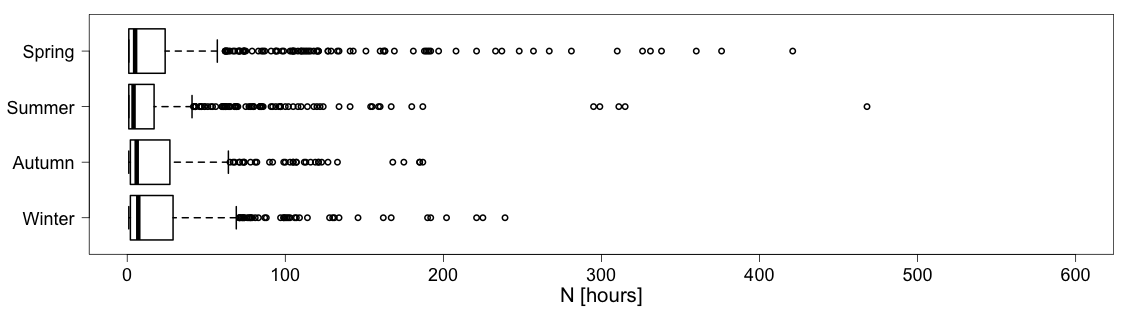}
		\caption{}
			\label{fig:dur0_boxplot}
	\end{subfigure} 
	
	\begin{subfigure}[t]{0.75\textwidth}
		\includegraphics[width=1\linewidth]{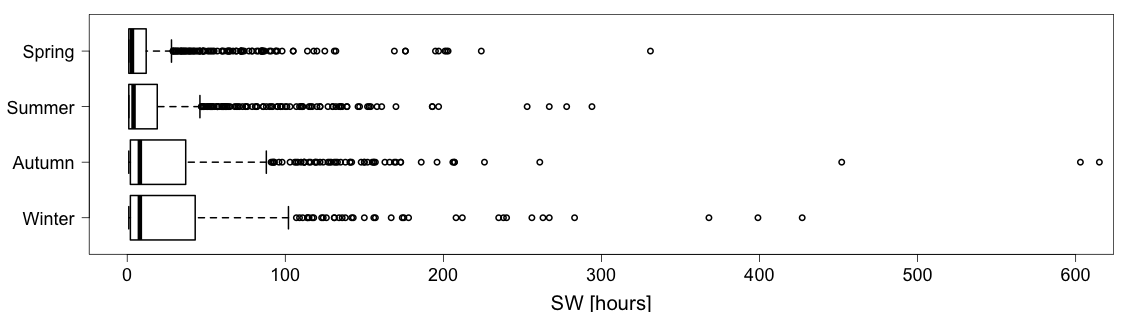}
		\caption{}
			\label{fig:dur1_boxplot}
	\end{subfigure}
	\caption{Box plots of the observed durations (a) $N$ and (b) $SW$ for the four seasons. }
	\label{fig:dur_boxplots}
 \end{figure}

We could not find adequate parametric univariate distributions among well-known families and suspect that this is caused by the many one-valued observations (i.e., many durations are $1$ hour). However, we did not investigate if and how so-called zero-inflated probability distributions could be adapted to this problem \cite[e.g.,][]{Zuur2009}. For simplicity, and since we do not need to extrapolate beyond the range of the observations, we used the empirical distribution functions. Figure~\ref{fig:dur_boxplots} shows box plots of the observations for $N$ and $SW$ for the four seasons. Seasonal differences are more pronounced for $SW$.

%The whiskers represent the lowest and highest simulated data point still within $1.5$ times the interquartile range from the lower and higher quartile, respectively. The open circles represent data points outside $1.5$ times the interquartile range from the lower and higher quartile, respectively.

\begin{figure}[t]
	\centering 
	\begin{subfigure}[t]{0.24\textwidth}
		\includegraphics[width=1\linewidth]{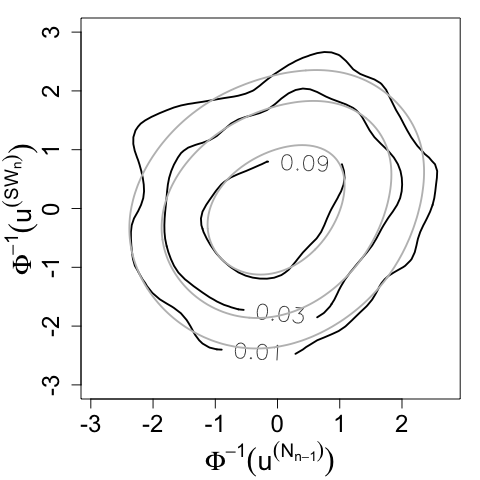}
		\caption{}
	\end{subfigure} 
	\begin{subfigure}[t]{0.24\textwidth}
		\includegraphics[width=1\linewidth]{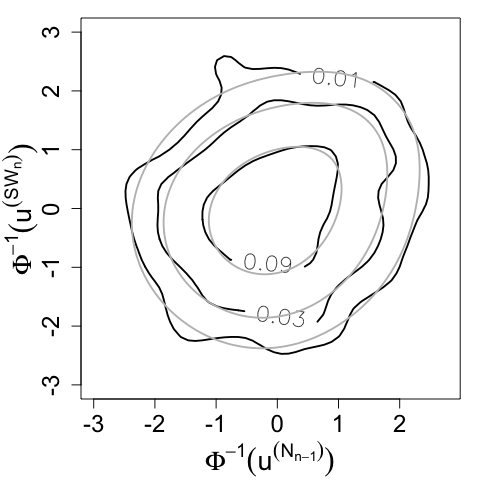}
		\caption{}
	\end{subfigure} 
	\begin{subfigure}[t]{0.24\textwidth}
		\includegraphics[width=1\linewidth]{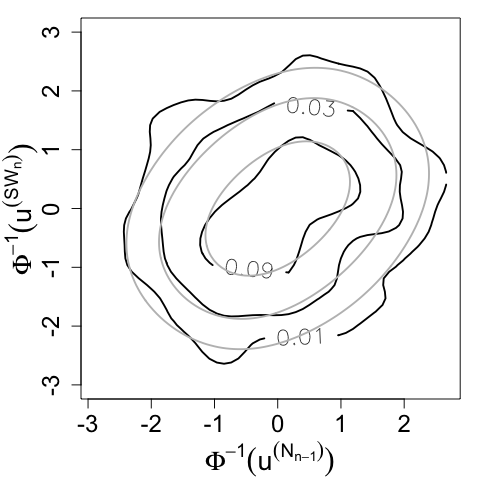}
		\caption{}
	\end{subfigure} 
	\begin{subfigure}[t]{0.24\textwidth}
		\includegraphics[width=1\linewidth]{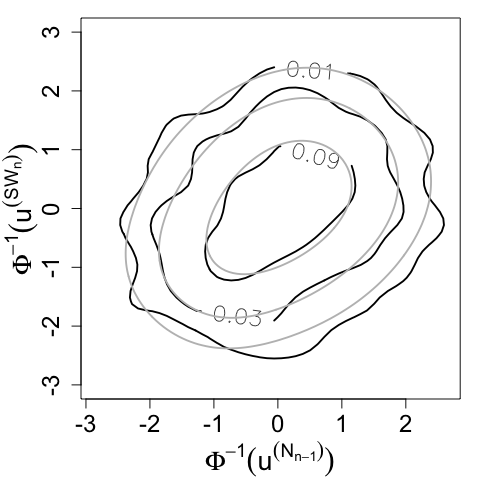}
		\caption{}
	\end{subfigure} 
	\caption{Bivariate density contours for observed (black) and simulated (gray) data of the pair $(N_{n-1},SW_n)$ for (a) spring, (b) summer, (c) autumn and (d) winter.}
	\label{fig:contours_01}
 \end{figure}

\begin{figure}[t]
	\centering 
	\begin{subfigure}[t]{0.24\textwidth}
		\includegraphics[width=1\linewidth]{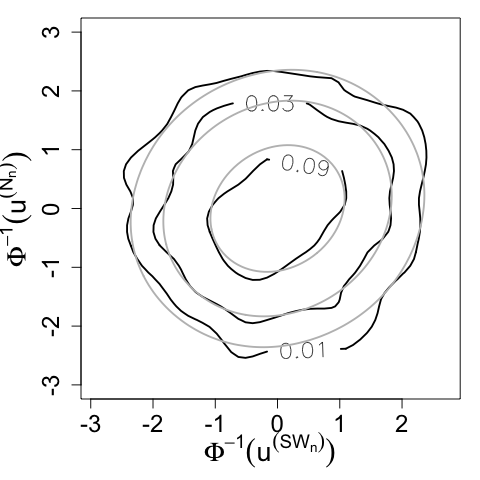}
		\caption{}
	\end{subfigure} 
	\begin{subfigure}[t]{0.24\textwidth}
		\includegraphics[width=1\linewidth]{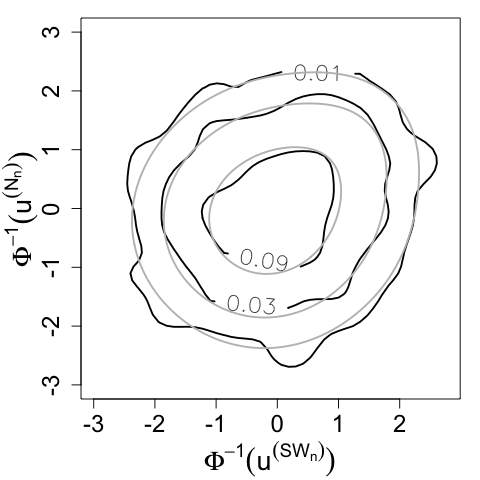}
		\caption{}
	\end{subfigure} 
	\begin{subfigure}[t]{0.24\textwidth}
		\includegraphics[width=1\linewidth]{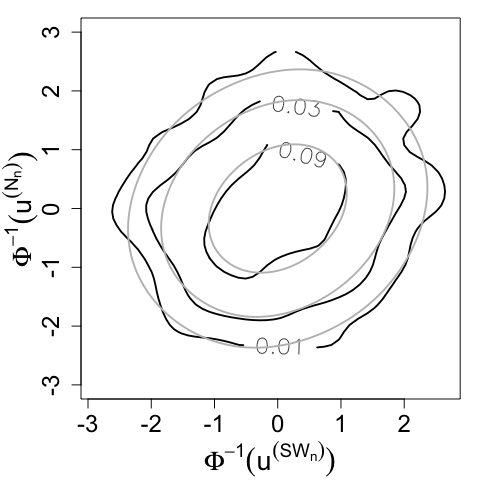}
		\caption{}
	\end{subfigure} 
	\begin{subfigure}[t]{0.24\textwidth}
		\includegraphics[width=1\linewidth]{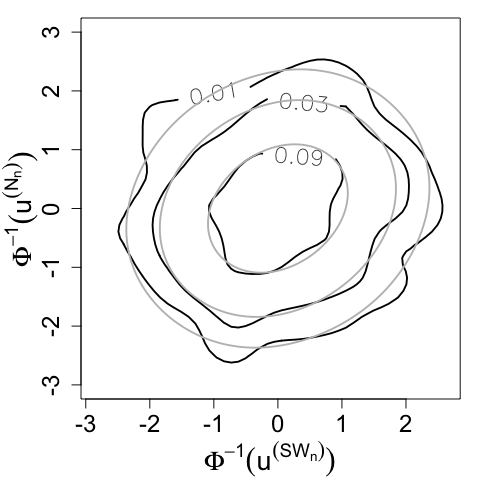}
		\caption{}
	\end{subfigure} 
	\caption{Bivariate density contours for observed (black) and simulated (gray) data of the pair $(SW_{n},N_n)$ for (a) spring, (b) summer, (c) autumn and (d) winter.}
	\label{fig:contours_10}
 \end{figure}

The last column of Table~\ref{tab:cops_renewal} shows the empirical values of Kendall's $\tau$ for $(N_{n-1},SW_n)$, $n \geq 2$, and for $(SW_{n},N_n)$, $n \geq 1$. Both variable pairs are positively associated. Thus, a wave direction regime tends to persist longer when the duration in the preceding regime was long than when it was short. This tendency is stronger and seasonal differences more distinct for $(N_{n-1},SW_n)$ than for $(SW_n,N_n)$. Bivariate copulas were selected according to the Akaike information criterion (AIC) using the VineCopula package \citep{Schepsmeier2017}, which compares twelve different families and, if applicable, their rotated versions. The corresponding parameters were estimated by maximum likelihoods. The selected families, parameter estimates and Kendall's $\tau$ for $(N_{n-1},SW_n)$, $n \geq 2$, and for $(SW_{n},N_n)$, $n \geq 1$, are also given in Table~\ref{tab:cops_renewal}. Bivariate density contour lines for observed data of the two variable pairs and the selected copula models indicate that the fit is valuable for our simulation purposes (Figures~\ref{fig:contours_01} and ~\ref{fig:contours_10}).

\begin{table}[thb]
\caption{Selected copula families, parameter estimates and Kendall's $\tau$ for the pairs $(N_{n-1},SW_n)$ and $(SW_n,N_n)$.}
\label{tab:cops_renewal}
\centering
\scriptsize
\begin{tabular} {c c c c c c c}
\toprule
Variable pair & Season & \multicolumn{4}{c}{Copula} & Empirical \\
& &  family &  par1 &  par2 &  $\tau$ &   $\tau$ \\ 
\midrule 
\multirow{4}{*}{$(N_{n-1},SW_n)$} & Spring & BB8 & $1.86$ & $0.68$ & $0.13$  &  $0.13$\\
& Summer& BB8 & $1.52$ & $0.81$ & $0.11$ & $0.11$ \\
& Autumn & Frank & $1.77$ & $-$ & $0.19$ & $0.20$\\
& Winter & Survival BB8 & $2.74$ & $0.59$ & $0.19$ & $0.18$\\
\midrule 
\multirow{4}{*}{$(SW_n,N_n)$} & Spring & Frank & $1.0.7$ & $-$ & $0.08$ & $0.08$ \\
& Summer& BB8 & $1.41$ & $0.85$ & $0.1$ & $0.09$ \\
& Autumn & Frank & $0.99$ & $-$ & $0.11$ & $0.11$ \\
& Winter & Frank & $1.02$ & $-$ & $0.11$  & $0.11$ \\
\bottomrule
\end{tabular}
\end{table} 

\subsection{Significant wave heights and mean zero-crossing periods}
\label{sec:wave-heights-and-periods}

\subsubsection{Modeling a limiting wave steepness condition} 
\label{sec:method-steepness}

\begin{figure}[t]
	\centering 
	\begin{subfigure}[t]{0.32\textwidth}
		\includegraphics[width=1\linewidth]{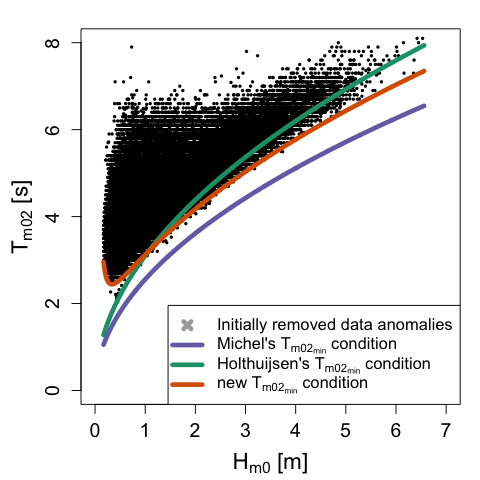}
		\caption{}
			\label{fig:scatter_hs_tp}
	\end{subfigure} 
	\begin{subfigure}[t]{0.32\textwidth}
		\includegraphics[width=1\linewidth]{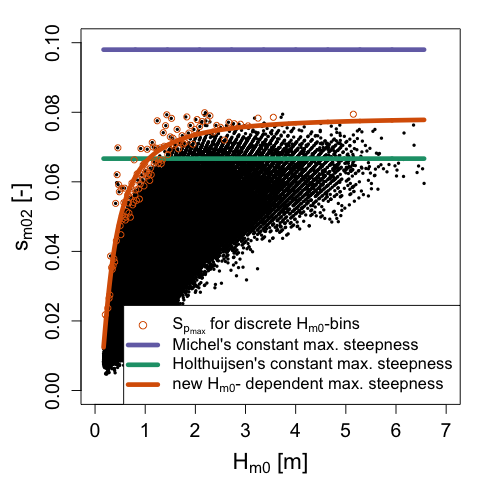}
		\caption{}
			\label{fig:scatter_hs_sp}
	\end{subfigure}
	\begin{subfigure}[t]{0.32\textwidth}
		\includegraphics[width=1\linewidth]{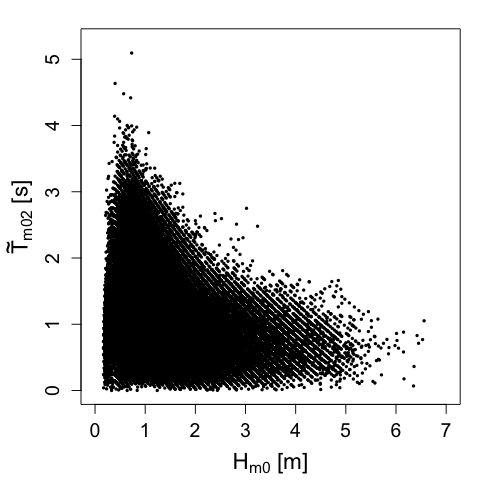}
		\caption{}
		\label{fig:scatter_hs_tp2}
	\end{subfigure}
	\caption{Scatter plots of (a) $H_{m0}$ and $T_{m02}$,  (b) $H_{m0}$ and $S_{m02}$, and (c) $H_{m0}$ and $\tilde{T}_{m02}$. Different proposed wave steepness limits are indicated in panels (a) and (b).}
 \end{figure}

While the relationship between $H_{m0}$ and $T_{m02}$ is in large part stochastic, there is a physical limit on the maximum steepness that individual waves can attain. As soon as waves approach this limit, they break\footnote{This is different in shallow water, where depth-induced breaking occurs before steepness-induced breaking.}. Wave steepness is defined as wave height divided by wave length, but can be formulated as a function of wave height and wave period. In terms of $H_{m0}$ and $T_{m02}$, it is \begin{equation} \label{eq:wave_steepness}
s_{m02} = \frac{2 \pi}{g} \frac{H_{m0}}{T_{m02}^2}.
\end{equation}
The limiting steepness condition is clearly visible in the scatter plot of $H_{m0}$ and $T_{m02}$ (Figure~\ref{fig:scatter_hs_tp}): For a given $T_{m02}$ the corresponding $H_{m0}$ cannot exceed a certain upper limit, or equivalently, for a given $H_{m0}$ the corresponding $T_{m02}$ is bounded from below. Nonetheless, we observed a few data points that appear to be unusually distant from the others (gray crosses in Figure~\ref{fig:scatter_hs_tp}). We suspect that these are anomalies in the measurements and substituted them by missing values before proceeding with the data analysis. 

Recent studies showed that bivariate distributions constructed with $3$- or $4$-parameter copula families can be suitable to reflect the limiting steepness condition and to represent the joint distribution of $H_{m0}$ and $T_{m02}$ \citep{vanem2016joint,jager2017joint,Zhang2018}. However, we cannot directly apply such approaches to the ARMA modeling in this study. Instead, we separate the deterministic part and the stochastic part of the relationship between $H_{m0}$ and $T_{m02}$ to model them individually. The idea is to remove the deterministic part by subtracting the lower bound from $T_{m02}$: \begin{equation}  \label{eq:TP_trafo}
\widetilde{T}_{m02} = T_{m02} - T_{m02_{min}}, 
\end{equation} where the lower bound can be determined from $H_{m0}$ and $s_{{m02}_{max}}$ as
\begin{equation}
T_{m02_{min}} = \sqrt{ \frac{2 \pi}{g}\frac{H_{m0}}{s_{{m02}_{max}}}}.
\end{equation} After this transformation, we proceed with time series modeling of the pair $( H_{m0}, \widetilde{T}_{m02})$.

A difficulty that arises is the choice of $s_{{m02}_{max}}$. Constant values for $s_{{m02}_{max}}$, as put forth by \citep{Michel1999,holthuijsen2010waves}, do not seem to fit our data set (cf. Figure~\ref{fig:scatter_hs_tp}). To achieve a better fit, we set out to estimate a maximum steepness condition from the data. The scatter plot indicates that observations of $s_{{m02}_{max}}$ are not constant, but depend on $H_{m0}$. More precisely, there appears to be a horizontal asymptote roughly below $s_{{m02}_{max}}=0.08$, while the observed $s_{{m02}_{max}}$ is rapidly decreasing for small $H_{m0}$. To account for this behavior we fit the curve \begin{equation}
s_{{m02}_{max}}(H_{m0}) = a \left( \frac{H_{m0}}{b} \right) ^{\left( \frac{c}{H_{m0}} \right)}, \quad a,b,c >0
\end{equation} to the data. 

Details on the motivation for this functional form and on the fitting procedure can be found in Appendix~\ref{app:wave-steepness}. The resulting estimates are $a=0.0782$, $b=999.4$cm (the upper bound was fixed at $1000$) and $c=7.674$cm. The fitted $s_{{m02}_{max}}$-curve is also shown in Figure~\ref{fig:scatter_hs_sp} and the corresponding $T_{m02_{min}}$-curve is depicted in Figure~\ref{fig:scatter_hs_tp} (orange lines). It should be noted that while this relation fits the observational data well in statistical sense, we do not propose to use this formulation to describe the physics of wave steepness-related breaking. 

Next, we substituted the data points that represented waves which are too steep according to the fitted limiting condition with missing values. (They amount to less than $0.08\%$ of the data.) Finally, we transformed $T_{m02}$ to $\widetilde{T}_{m02}$ (equation~\ref{eq:TP_trafo}). Figure~\ref{fig:scatter_hs_tp2} shows a scatter plot of the pair $(H_{m0},\widetilde{T}_{m02})$. In the remainder, we will denote $X^{(H_{m0})} = H_{m0}$, $X^{(\widetilde{T}_{m02} )} = \tilde{T}_{m02}$ to simplify notation.

\subsubsection{Normalization and Deseasonalization of data}
\label{sec:method-normalization-deseasonalization}

Neither of the processes $\lbrace X^{(H_{m0})}_t \rbrace_t$ and $\lbrace X^{(\tilde{T}_{m02})}_t  \rbrace_t $ is stationary. Inspection of the data suggests strong seasonal behavior (cf. Figures~\ref{fig:ts_x_hs} and ~\ref{fig:ts_x_tp}). Trends do not seem apparent. Moreover, the data are notably right skewed and strictly positive. Different methods can be used to analyze and model such time series data \citep[e.g.,][]{Box2015}. The prevailing approach is to deseasonalize the data and to model the seasonal components and the stationary component separately. 

We identified two main procedures for deseasonalization in the literature on wave parameter modeling and simulation that appeared successful. Suppose $x_t$, $t = 1,...,T$, is a time series for an arbitrary wave parameter. The first procedure involves two steps \citep{Stefanakos2006}. One step is to transform the data to reduce skewness:\begin{equation}
y_t = f(x_t), \qquad t=1,..., T,
\end{equation} with $f$ being a suitable monotone transformation function. 
This facilitates finding a adequate residual distribution of the ARMA model for the stationary component, but is also important for simulations, as will be explained in the next paragraph. A second step is to represent the transformed time series data as a realization of the following process: \begin{equation} \label{eq:decomposition}
y_t = \mu_t +\sigma_t z_t, \qquad t=1,..., T,
\end{equation} where $\mu_t $ and $\sigma_t $ are slowly changing non-stationary components exhibiting seasonal feature and $z_t$ is a high-frequency, stationary component. Each of the components is then modeled and $y_t$ is obtained by combining them according to equation~(\ref{eq:decomposition}). \cite{GuedesSoares1996} followed a similar procedure, however they first decomposed the data in seasonal and stationary components and then applied a skewness reducing transformation to the stationary data series. The second procedure relies on using a non-stationary distribution function to transform the data to standard normal \citep{Solari2011}. With this approach no additional transformation is necessary.

%ask claudia and thomas about drawbacks on the second method.
We used a method in line with the first approach to develop the simulation model.  As potential transformations we considered the Box-Cox family \citep{Box1964} \begin{equation}
f(x, \lambda) = \begin{cases}
\frac{x^\lambda -1}{\lambda}, \quad \lambda \not=0, \\
\log(x), \quad \lambda =0
\end{cases},
\end{equation} shifted logarithms \begin{equation}
f(x, c) = \log(x+c), \quad c\geq 0,
\end{equation}
and a transformation to standard normal using the empirical distribution function in the probability integral transform: \begin{equation}\label{eq:PIT}
f(x) = \Phi^{-1}(\hat{F}_{n}(x))
\end{equation} where $\Phi^{-1}$ is the inverse of the standard normal cumulative distribution function and $\hat{F}_n$ is the empirical distribution function estimated for the data $x_1, ..., x_T$. All three transformations have proven valuable in similar applications \citep{GuedesSoares1996,Cunha1999,Stefanakos2006}. 

The choice of transformation strongly influenced the simulation results that would be obtained at later stages. Box-Cox transformations whose $\lambda$ was estimated by maximum likelihood or a (not shifted) logarithm resulted in unrealistically high simulated values. For instance, simulated $H_{m0}$ were in the order of $30$m, while the highest observed is below $7$m. In contrast, choosing higher values of $\lambda$ or shifting the logarithm by a positive constant $c$ would result in reasonable maximum heights, but also in negative ones. These issues arose both when transforming the data before the decomposition as well as when transforming the stationary component. 

\begin{figure}[thb]
	\centering 
	\begin{subfigure}[t]{0.32\textwidth}
		\includegraphics[width=1\linewidth]{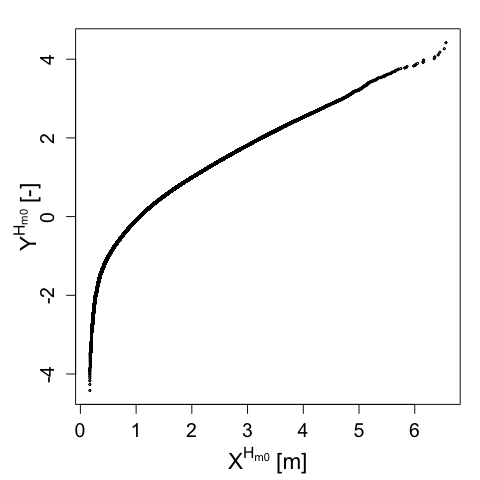}
		\caption{}
			\label{fig:hs_normalization}
	\end{subfigure} 
	\begin{subfigure}[t]{0.32\textwidth}
		\includegraphics[width=1\linewidth]{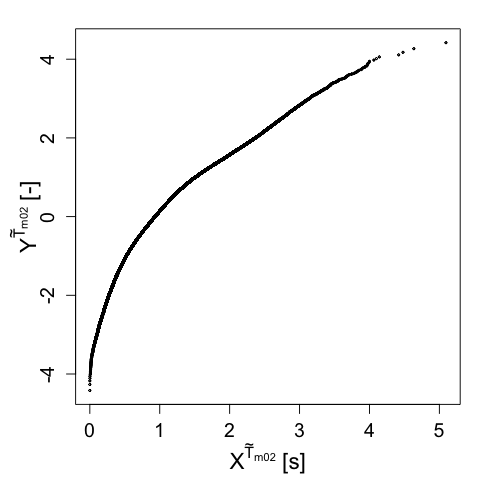}
		\caption{}
			\label{fig:tp2_normalization}
	\end{subfigure}
		\caption{Normalizing transformation for (a) $H_{m0}$ and (b) $T_{m02}$}
	\label{fig:normalization}
 \end{figure}

The transformation to standard normal via the probability integral transform resulted in simulation values that were representative of real values, at least when applied before the decomposition. When transforming after the decomposition, the simulated values of $\mu_t$, $\sigma_t$ and $Z_t$ would sometimes combine to a negative value. This is not surprising, since we modeled $\mu_t$ and $\sigma_t$ independent of $Z_t$, as will be explained in the next sections. Hence, the transformation to standard normal via the probability integral transform before a decomposition was chosen. Figures~\ref{fig:hs_normalization} and~\ref{fig:tp2_normalization} show the relationship between transformed and original variables.

 \begin{figure}[thb]
 	\centering 
 	\begin{subfigure}[t]{0.49\textwidth}
 		\includegraphics[width=1\linewidth]{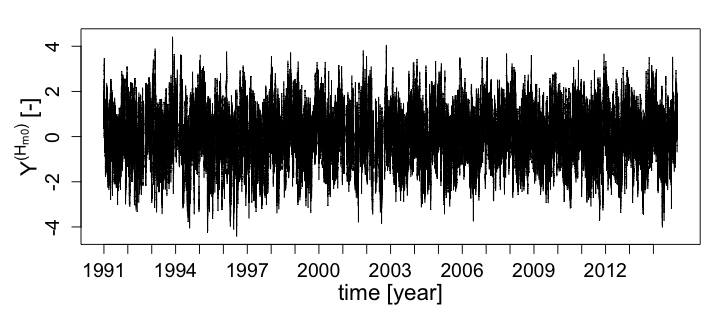}
 		\caption{}
 			\label{fig:ts_y_hs}
 	\end{subfigure} 
 	\begin{subfigure}[t]{0.49\textwidth}
 		\includegraphics[width=1\linewidth]{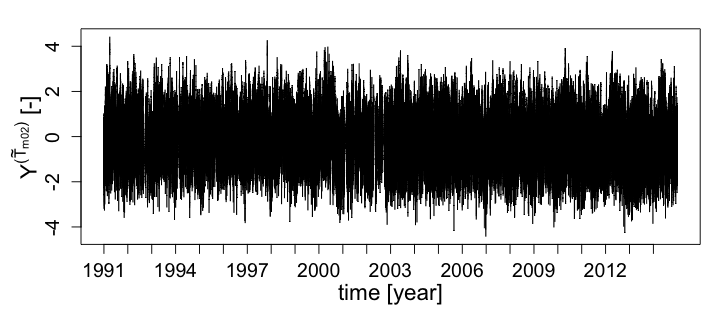}
 		\caption{}
 			\label{fig:ts_y_tp}
 	\end{subfigure}
 	\caption{Time series of normalized (a) $H_{m0}$ and (b) $T_{m02}$ from January 1989 to December 2014.}
 \end{figure}

After each of the time series $x_t$, $t=1,...,T$, has been transformed to the series $y_t$, the seasonal components $\mu_t$ and $\sigma_t$ of equation~\ref{eq:decomposition} are extracted using a smoothing technique. $\mu_t$ can be interpreted as a local mean and $\sigma_t$ as a local standard deviation. Both were computed with sliding windows and an Epanechnikov kernel as weighting function for smoothing, as follows: \begin{equation}
\mu_t = \frac{1}{2t'} \sum_{k=t-t'}^{t+t'} K_{2t'}(x_t-x_k)
\end{equation} and
\begin{equation}
\sigma_t = \sqrt{ \frac{1}{2t'} \sum_{k=t-t'}^{t+t'}  K_{2t'}((x_t-\mu_t)^2-(x_k-\mu_k)^2)},
\end{equation}
where $K_{2k}$ is the Epanechnikov (parabolic) kernel. The bandwidth was set to $720$ hours, which amounts to 30 days and is in line with the common practice to deseasonalize oceanographic variables via monthly statistics \citep[recent examples are][]{wahl2016probabilistic,davies2017Improved}. 

%The computed local mean time series, $\mu_t^{(H_{m0})}$ and $\mu_t^{(\widetilde{T}_{m02})}$, are shown in Figures~\ref{fig:ts_mu_hs} and~\ref{fig:ts_mu_tp}. The corresponding local standard deviation time series, $\sigma_t^{(H_{m0})}$ and $\sigma_t^{(\widetilde{T}_{m02})}$, are shown in Figures~\ref{fig:ts_sig_hs} and~\ref{fig:ts_sig_tp}. Finally, the resulting stationary components, $Z_t^{(H_{m0})}$ and $Z_t^{(\widetilde{T}_{m02})}$, are shown in Figures~\ref{fig:ts_z_hs} and~\ref{fig:ts_z_tp}. 

\subsubsection{Model for the non-stationary components}
\label{sec:method-nonstationary}

\begin{figure}[thb]
 	\centering 
 	\begin{subfigure}[t]{0.49\textwidth}
 		\includegraphics[width=1\linewidth]{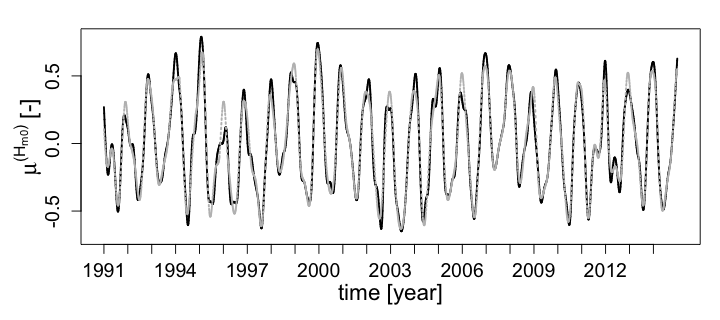}
 		\caption{}
 			\label{fig:ts_mu_hs}
 	\end{subfigure} 
 	\begin{subfigure}[t]{0.49\textwidth}
 		\includegraphics[width=1\linewidth]{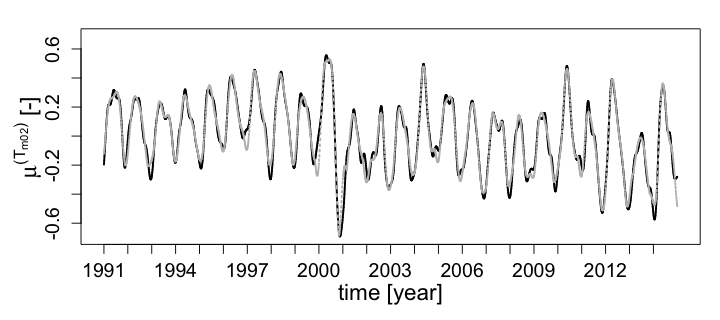}
 		\caption{}
 			\label{fig:ts_mu_tp}
 	\end{subfigure}

 	\centering 
 	\begin{subfigure}[t]{0.49\textwidth}
 		\includegraphics[width=1\linewidth]{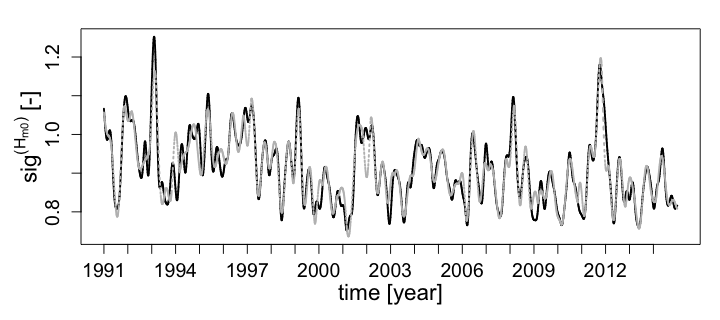}
 		\caption{}
 			\label{fig:ts_sig_hs}
 	\end{subfigure} 
 	\begin{subfigure}[t]{0.49\textwidth}
 		\includegraphics[width=1\linewidth]{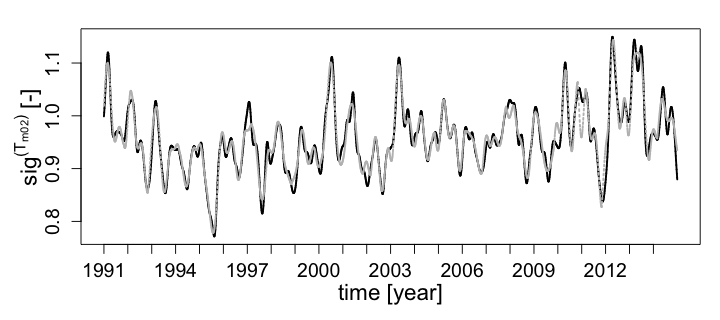}
 		\caption{}
 			\label{fig:ts_sig_tp}
 	\end{subfigure}
 	\caption{Seasonal mean and standard deviation processes processes: (a) $\mu^{(H_{m0})}_t$, (b) $\mu^{(T_{m02})}_t$, (c) (a) $\sigma^{(H_{m0})}_t$, and (d) (a) $\sigma^{(T_{m02})}_t$.}
  \end{figure}

Figures ~\ref{fig:ts_mu_hs}-\ref{fig:ts_sig_tp} show the seasonal patterns and the inter-year variability of the non-stationary processes. At this stage, we neither studied climatological covariates nor cycles longer than $1$ year, which is different to many of the studies reviewed in the introduction. Instead, we developed another approach assuming that it is more important to represent the range of inter-year differences than their temporal sequencing. 

We start our modeling by following earlier studies \cite{Mendez2006,Mendez2007,serafin2014simulating} in using  a $5$-factor Fourier series in order to represent both annual and semiannual cycles following earlier studies. However, we additionally introduce randomness in the Fourier coefficients, which will cause them to vary each year and thus produce inter-year variations is simulated time series of $H_{m0}$ and $T_{m02}$.

For simplicity, we let the coefficients vary randomly according to a multivariate Gaussian distribution. This also accounts for dependencies between the four local mean and standard deviation processes. In the future,  it might be worthwhile to further investigate, if the coefficients could be predicted by climatic covariates. 

The first step of this approach involves dissecting the $24$-year data series $\mu_t^{(H_{m0})}$, $\mu_t^{(\widetilde{T}_{m02})}$, $\sigma_t^{(H_{m0})}$ and $\sigma_t^{(\widetilde{T}_{m02})}$ into $1$-year segments. To easily identify them, we re-index the series using a double index, $k=1,...,24$ for the year and $\tau= 1,...,8766$ for the hour within the year\footnote{8766 hours correspond to one year when``correcting" for leap years.}. Then, we estimate $5$-factor Fourier series for each segment:\begin{equation}
f^{(\mu^{(i)}_k)}(\tau) = a_0^{(\mu^{(i)}_k)} +  a_{1}^{(\mu^{(i)}_k)} \cos \left( \frac{2 \pi \tau}{T} \right)  + a_{2}^{(\mu^{(i)}_k)}  \sin \left(  \frac{2\pi \tau}{T} \right) +  a_{3}^{(\mu^{(i)}_k)} \cos \left(  \frac{4 \pi \tau}{T} \right)  + a_{4}^{(\mu^{(i)}_k)}  \sin \left(  \frac{4\pi \tau}{T} \right) 
\end{equation}
and
\begin{equation}
f^{(\sigma^{(i)}_k)}(\tau) = a_0^{(\sigma^{(i)}_k)} +  a_{1}^{(\sigma^{(i)}_k)} \cos \left( \frac{2 \pi \tau}{T} \right)  + a_{2}^{(\sigma^{(i)}_k)}  \sin \left( \frac{2\pi \tau}{T} \right)  +  a_{3}^{(\sigma^{(i)}_k)} \cos \left( \frac{4 \pi \tau}{T} \right) + a_{4}^{(\sigma^{(i)}_k)}  \sin \left( \frac{4\pi \tau}{T} \right) 
\end{equation} where $T=8766$ and $i= \lbrace H_{m0}, \widetilde{T}_{m02} \rbrace$. To obtain the $24$-year series, the fitted $1$-year segments are concatenated, for example $f^{(\mu^{(i)})} = \left[f^{(\mu^{(i)}_1)}, ..., f^{(\mu^{(i)}_{24})} \right]$, and discontinuities at the transitions from one year to another are smoothed out using cubic spline interpolation.  The continuous  fitted series obtained in this way explain more than $90\%$ of the variance in the corresponding data series (Table~\ref{tab:r-squared}).

\begin{table}[htb]
\caption{Coefficient of determination $R^2$ for the fitted seasonal mean and standard deviation series.}
\label{tab:r-squared}
\centering
\begin{tabular}{c c c c}
  \toprule
$\mu_t^{(H_{m0})}$ & $\mu_t^{(\widetilde{T}_{m02})}$ & $\sigma_t^{(H_{m0})}$ &$\sigma_t^{(\widetilde{T}_{m02})}$ \\
  \midrule
$0.96$ & $0.91$ & $0.98$ & $0.94$ \\
   \bottomrule
\end{tabular}
\end{table}

In the next step, we assume that the estimated coefficients $a_m^{(i,k)}$ are i.i.d. observations of random variables $A_m^{(i)}$, $m=0,...,4$, and estimate a joint distribution for them. Since the sample size is small ($N=24$) compared to the dimension of the random vector ($d=20$), an extensive analysis of its distribution seems infeasible and for simplicity we assumed it to be multivariate Gaussian. 

We modeled the distribution in terms of univariate Gaussian marginals and a multivariate Gaussian copula. Despite the relatively high dimensionality, the correlation matrix is sparse. On one hand, the basis functions of a Fourier series are mutually orthogonal, hence their coefficients uncorrelated leading to many zero-valued entries. On the other hand, we performed the bivariate asymptotic independence test based on Kendall's $\tau$ for the remaining pairs of coefficients, which is implemented in the VineCopula package \citep{Schepsmeier2017}. According to the test, most pairs are independent. Only three correlations were found to be significant: $\rho(A_0^{(\sigma^{(H_{m0})})},A_0^{(\sigma^{(\widetilde{T}_{m02})})}) = -0.7$, $\rho(A_1^{(\mu^{(\widetilde{T}_{m02})})},A_1^{(\sigma^{(\widetilde{T}_{m02})})}) = 0.6$, and $\rho(A_2^{(\mu^{(H_{m0})})},A_2^{(\mu^{(\widetilde{T}_{m02})})}) = -0.5$, where $\rho$ denotes the product moment correlations. Thus, these correlation values correspond to the only three non-zero off-diagonal elements of the the correlation matrix parameterizing the Gaussian copula. The parameters of the univariate distributions were estimated by maximum likelihood and can be found in Table~\ref{tab:local-means-margins}.

\begin{table}[h!]
\caption{Parameters (mean, standard deviation) of the univariate normal distributions of the Fourier coefficients}
\label{tab:local-means-margins}
\centering
\scriptsize
\begin{tabularx} {\textwidth} {l | Y Y Y Y Y}
\toprule
& $A_0$ & $A_1$ & $A_2$ & $A_3$ & $A_4$ \\
\midrule 
$\mu^{(H_{m0})}$ & $ (0, 0.09)$ & $ (0.39 , 0.09)$ & $(-0.10,0.12)$ &  $(0.03,0.08)$ & $(-0.03 ,0.10)$ \\
$\mu^{(\widetilde{T}_{m02})}$ & $( 0, 0.10)$ & $(-0.21 , 0.06)$ & $(0.14,0.1)$ & $(-0.03, 0.04)$ &  $(0, 0.07)$\\
$\sigma^{(H_{m0})}$ & $( 0.91, 0.05)$ & $(0.01 , 0.05)$ & $(0,0.06)$ & $(-0.01, 0.03)$ &  $(0.01, 0.05)$\\
$\sigma^{(\widetilde{T}_{m02})}$ & $( 0.96, 0.04)$ & $(0 , 0.04)$ & $(0.03,0.03)$ & $(-0.01, 0.02)$ &  $(0, 0.03)$\\
\bottomrule
\end{tabularx}
\end{table}

%\begin{figure}[thb]
%	\centering 
%	\begin{subfigure}[b]{0.49\textwidth}
%		\includegraphics[width=1\linewidth]{figures/sim_mu_hs.png}
%		\caption{}
%	\end{subfigure}
%	\begin{subfigure}[b]{0.49\textwidth}
%		\includegraphics[width=1\linewidth]{figures/sim_mu_tp2.png}
%		\caption{}
%	\end{subfigure}
%	
%	\begin{subfigure}[b]{0.49\textwidth}
%		\includegraphics[width=1\linewidth]{figures/sim_sig_hs.png}
%		\caption{}
%	\end{subfigure}
%	\begin{subfigure}[b]{0.49\textwidth}
%		\includegraphics[width=1\linewidth]{figures/sim_sig_tp2.png}
%		\caption{}
%	\end{subfigure}
%	\caption{24 observed years (black) and 500 simulated years (gray)}
%\end{figure}

\subsubsection{Model for the stationary components}
\label{sec:method-stationary}

The deseasonalised processes, obtained through \begin{equation}
z_t^{(i)} = \frac{y_t^{(i)} - \mu_t^{(i)} }{\sigma_t^{(i)} },
\end{equation} 
 are represented as ARMA. To find adequate orders $p$ and $q$ for the two series, we followed this process: We used well known properties of the behavior of the ACF and PACF to make an initial guess about the orders p and q (for methodological background on ARMA processes see~\ref{sec:background-arma}). For these p and q, we estimated the parameters of the ARMA based on maximum likelihood, as implemented in the arima() function of R's stats package \citep{CoreTeam2015}.  With diagnostic plots, we checked, if the obtained residual series resembled white noise and if the model could produce simulated series with ACF and PACF behavior resembling the observed. If this was not the case, we iteratively increased the orders $p$ and $q$, estimated model parameters and reassessed the adequacy of the model. 
 
\begin{figure}[thb]
 	\centering 
 	\begin{subfigure}[t]{0.2\textwidth}
 		\includegraphics[width=1\linewidth]{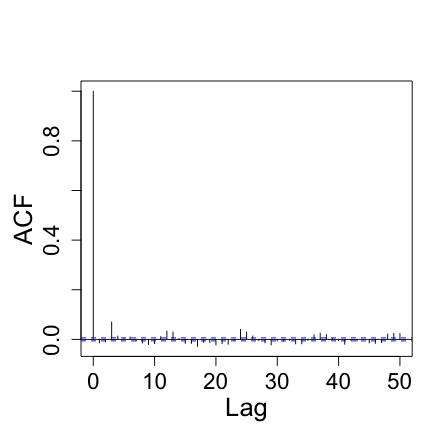}
 		\caption{}
 	\end{subfigure} 
 	\begin{subfigure}[t]{0.2\textwidth}
 		\includegraphics[width=1\linewidth]{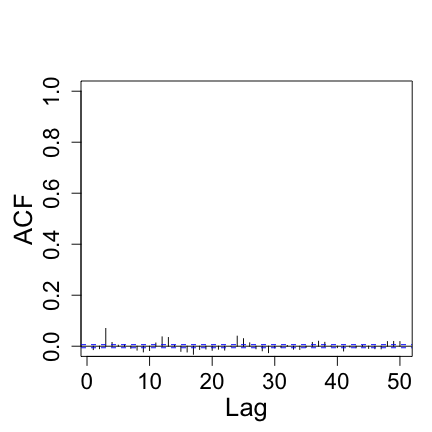}
 		\caption{}
 	\end{subfigure}
   	\begin{subfigure}[t]{0.2\textwidth}
   		\includegraphics[width=1\linewidth]{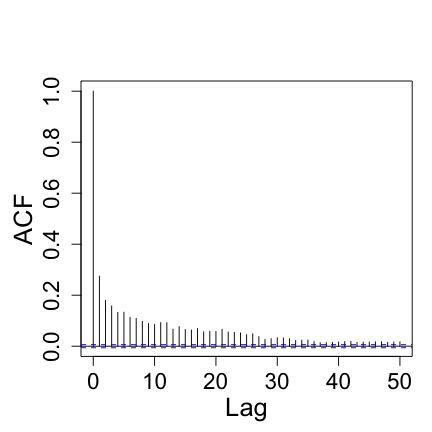}
   		\caption{}
   	\end{subfigure} 
   	
   	\begin{subfigure}[t]{0.2\textwidth}
   		\includegraphics[width=1\linewidth]{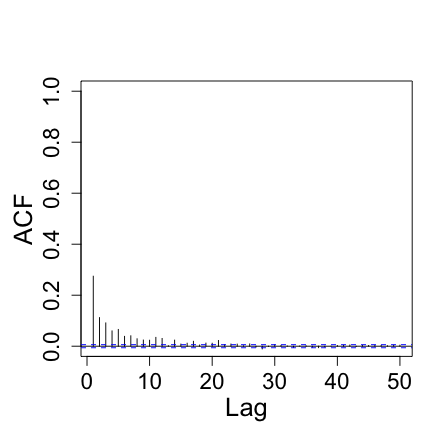}
   		\caption{}
   	\end{subfigure}
    	\begin{subfigure}[t]{0.2\textwidth}
    		\includegraphics[width=1\linewidth]{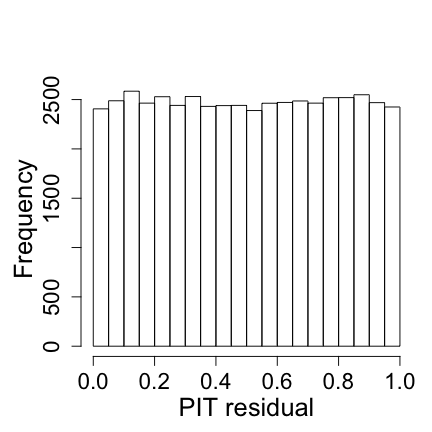}
    		\caption{}
    	\end{subfigure} 
    	\begin{subfigure}[t]{0.2\textwidth}
    		\includegraphics[width=1\linewidth]{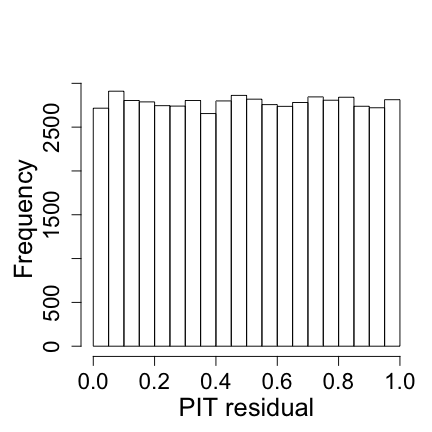}
    		\caption{}
    	\end{subfigure}
    \caption{Diagnostic plots for $\epsilon_t^{(H_{m0})}$: (a) ACF of residuals, (b) PACF of residuals, (c) ACF of squared residuals, (d) PACF of squared residuals, (e) histogram of PIT residuals for northern directions, (f) histogram of PIT residuals for southern directions.}
    \label{fig:diagnostic_Hm0_residuals}
\end{figure}

\begin{figure}[thb]
  	\centering 
  	\begin{subfigure}[t]{0.2\textwidth}
  		\includegraphics[width=1\linewidth]{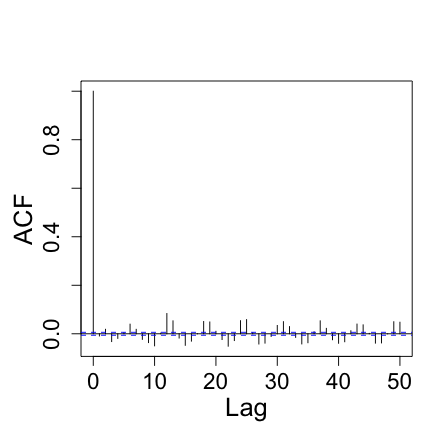}
  		\caption{}
  	\end{subfigure} 
  	\begin{subfigure}[t]{0.2\textwidth}
  		\includegraphics[width=1\linewidth]{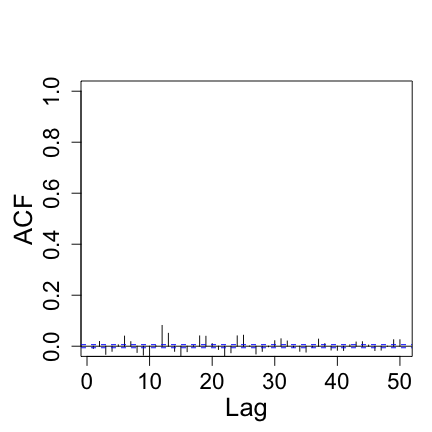}
  		\caption{}
  	\end{subfigure}
    	\begin{subfigure}[t]{0.2\textwidth}
    		\includegraphics[width=1\linewidth]{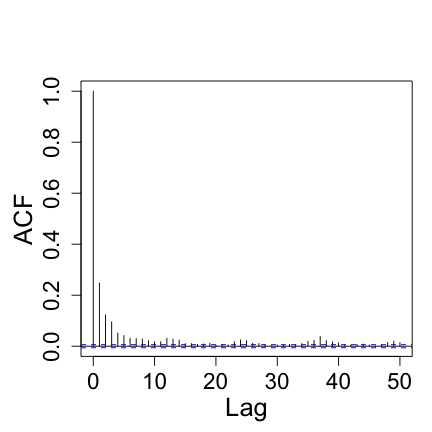}
    		\caption{}
    	\end{subfigure} 
    	
    	\begin{subfigure}[t]{0.2\textwidth}
    		\includegraphics[width=1\linewidth]{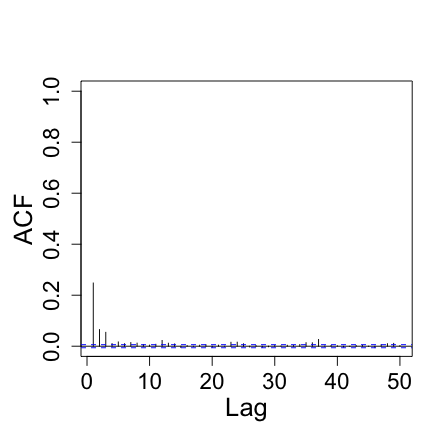}
    		\caption{}
    	\end{subfigure}
    	\begin{subfigure}[t]{0.2\textwidth}
    		\includegraphics[width=1\linewidth]{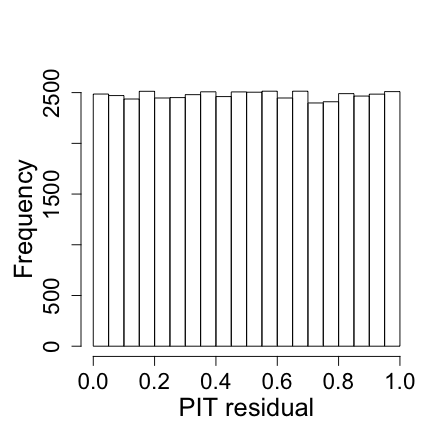}
    		\caption{}
    	\end{subfigure} 
    	\begin{subfigure}[t]{0.2\textwidth}
    		\includegraphics[width=1\linewidth]{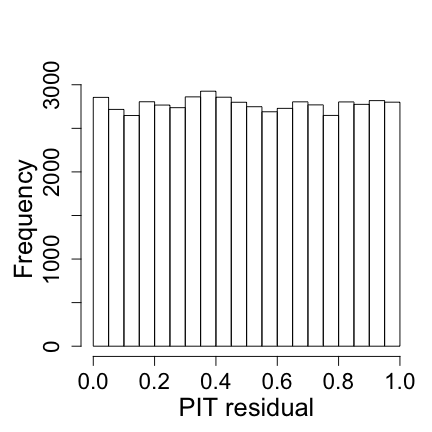}
    		\caption{}
    	\end{subfigure}
 	\caption{Diagnostic plots for $\epsilon_t^{(T_{m02})}$: (a) ACF of residuals, (b) PACF of residuals, (c) ACF of squared residuals, (d) PACF of squared residuals, (e) histogram of fitted quantiles for northern directions, (f) histogram of fitted quantiles for southern directions.}
 	\label{fig:diagnostic_Tm02_residuals}
  \end{figure}

The initial guess for $z_t^{(H_{m0})}$ was $p=3$ and $q=0$, since its ACF decays slowly and its PACF has a cut of at lag $3$. This choice of parameters was found to be adequate, which we show here with the plots of the ACF and PACF of the obtained residuals and squared residuals in Figures~\ref{fig:diagnostic_Hm0_residuals}(a)-(d). The initial guess for $z_t^{(T_{m02})}$ was $p=2$ and $q=2$, since both its ACF and PACF decay slowly while being dominated by damped sine waves. These orders were not found to be adequate yet, but $p=3$ and $q=2$ were. The plots of the ACF and PACF of the obtained residuals and squared residuals are given in Figures~\ref{fig:diagnostic_Tm02_residuals}(a)-(d). The estimated parameters are listed in Table~\ref{tab:armas}. 

  \begin{figure}[thb]
 	\centering 
 	\begin{subfigure}[t]{0.49\textwidth}
 		\includegraphics[width=1\linewidth]{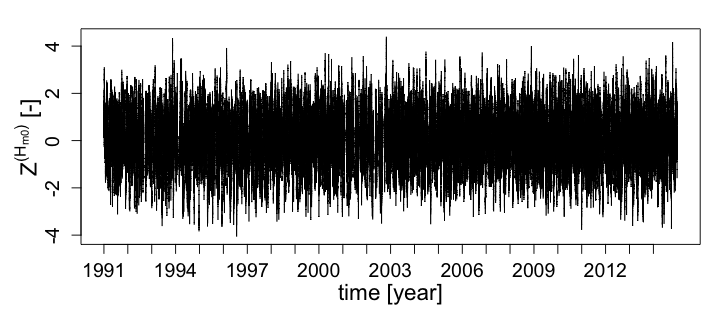}
 		\caption{}
 			\label{fig:ts_z_hs}
 	\end{subfigure} 
 	\begin{subfigure}[t]{0.49\textwidth}
 		\includegraphics[width=1\linewidth]{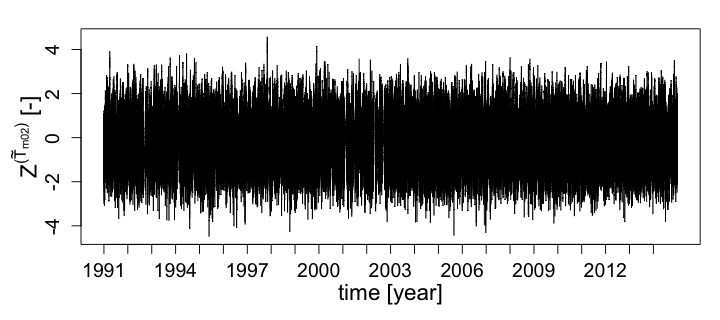}
 		\caption{}
 			\label{fig:ts_z_tp}
 	\end{subfigure}
 		\caption{Stationary components of normalized (a) $H_{m0}$ and (b) $T_{m02}$ time series as obtained after the decomposition.}
 	\label{fig:time_series}
  \end{figure}

The residuals, $\epsilon_t^{(i)}$, of the fitted $ARMA(p,q)$ model for $z_t^{(i)}$ should be (almost) i.i.d. and following an arbitrary distribution with zero mean and constant variance. Because we expect differences across the wave direction regimes, we estimate conditional distributions for the residuals given the regime, $\mathcal{E}^{(i)} | R^\Theta = k$, for each $k \in \lbrace 0,1 \rbrace$. We cannot expect these distributions to be Gaussian and we need to find a suitable parametric family of distributions. We considered the following families in this study: normal, skew-normal, t and  skew-t. We select the family and attest its goodness of fit with visual diagnostic tools. We examine qq-plots and check if applying a probability integral transform (PIT) to the residuals via the selected family results in uniformly distributed values. We refer to these as PIT residuals and visually assess their uniformity with histograms. All four univariate distributions are best approximated by a skew-t family. The estimated parameters can be found in Table~\ref{tab:residual-dists}. Diagnostic plots to verify the adequacy the marginal distribution are shown in Figures~\ref{fig:diagnostic_Hm0_residuals}(e) and (f)  and~\ref{fig:diagnostic_Tm02_residuals} (e) and (f).

\begin{table}[thb]
\caption{Coefficients of ARMA models for $Z^{(H_{m0})}_t$ and $Z^{(T_{m02})}_t$. Standard errors are given in parenthesis.}
\label{tab:armas}
\centering
\scriptsize
\begin{tabular} {c c c c c c c}
\toprule
Process & ar1 & ar2 &ar3 & ma1 & ma2 & intercept \\
\midrule
 $Z^{(H_{m0})}_t$ & $1.07$ $(0.00)$ &  $0.10$ $(0.01)$ & $-0.18$ $(0.00)$ & - & - &   $0.00$ $(0.03)$\\
  $Z^{(\widetilde{T}_{m02})}_t$ & $2.63$ $(0.00)$ &  $-2.54$ $(0.00)$ &  $0.89$ $(0.00)$ &  $-1.62$ $(0.00)$ & $0.83$ $(0.00$) &  $0.00$ $(0.01)$\\
\bottomrule
\end{tabular}
\normalsize
\end{table}

\begin{table}[thb]
\caption{Parameters (mu, sigma, skew, shape) of regime-dependent skew-t distributions for residuals $\mathcal{E}^{(H_{m0})}$ and $\mathcal{E}^{(T_{m02})}$.}
\label{tab:residual-dists}
\centering
\scriptsize
\begin{tabular} {c c c }
\toprule
$k$ &$\hat{F}_{\mathcal{E}^{(H_{m0})} \mid R^\Theta=k}$  & $\hat{F}_{\mathcal{E}^{(T_{m02})} \mid R^\Theta=k}$   \\
\midrule
0 & $(-0.01,0.16,1.07, 4.80)$ & $(0.05, 0.35, 0.87, 5.58)$ \\
1 & $(0.01, 0.17, 1.13, 5.36)$ & $(-0.04, 0.43, 0.90, 5.52)$ \\
\bottomrule
\end{tabular}
\normalsize
\end{table} 

\begin{figure}[thb]
	\centering
    	\begin{subfigure}[t]{0.32\textwidth}
    		\includegraphics[width=1\linewidth]{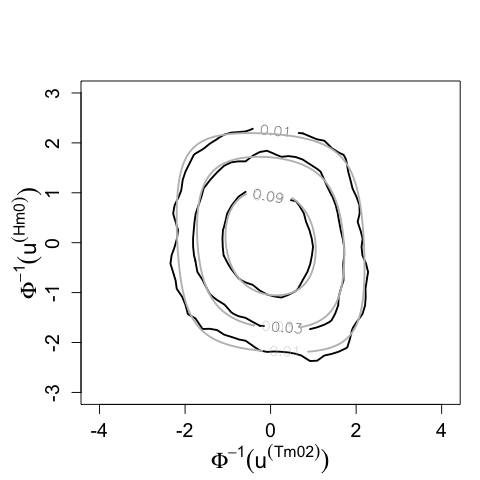}
    		\caption{}
    	\end{subfigure} 
    	\begin{subfigure}[t]{0.32\textwidth}
    		\includegraphics[width=1\linewidth]{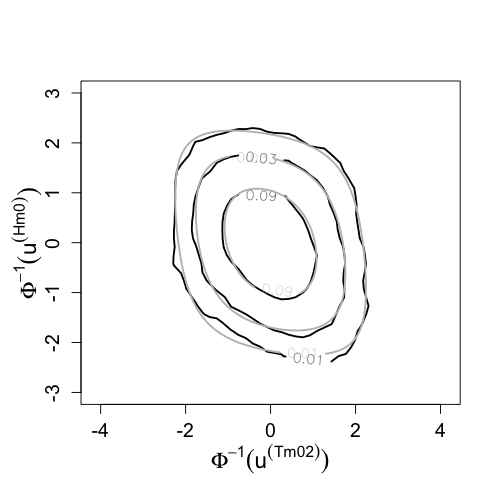}
    		\caption{}
    	\end{subfigure}
 	\caption{Bivariate density contour plots for residuals $\mathcal{E}^{(H_{m0})}$ and $\mathcal{E}^{(T_{m02})}$ (a) corresponding to northern wave directions and (b) to south-western wave directions.}
 	\label{fig:contours_residuals}
  \end{figure}

Finally, the residual processes, $\epsilon^{(1)}_t$ and $\epsilon^{(2)}_t$ could depend on each other, because the original time series are interrelated. Therefore, we construct regime-dependent bivariate residual distributions via copulas.  We fit them on the empirical ranks of the conditional residuals normalized to $(0,1)$. As in section~\ref{sec:method_WA}, we use the AIC criteria for model selection and estimate the parameters by maximum likelihood. Table~\ref{tab:residual-vine} contains the selected bi-variate copula families, estimated parameters and Kendall's $\tau$. Figure~\ref{fig:contours_residuals} shows the bivariate density contours for observed and simulated residuals and attests, by visual diagnostic, a good fit of the selected copula models.

\begin{table}[thb]
\caption{Bi-variate copula families, their parameters and Kendall's $\tau$ for regime-dependent copulas for residuals $\mathcal{E}^{(H_{m0})}$ and $\mathcal{E}^{(T_{m02})}$.}
\label{tab:residual-vine}
\centering
\scriptsize
\begin{tabular} {c | c c c c }
\toprule
 &
 \multicolumn{4}{c}{$\mathcal{E}^{(H_{m0})},\mathcal{E}^{(T_{m02})} \mid R=k$} \\
$k$ & family & par & par2 & tau \\
\midrule
$0$ & $t$ & $-0.09$ & $ 5.53$ & $-0.06$\\
$1$ & $t$ & $-0.23$ & $6.36$ & $-0.14$ \\
\bottomrule
\end{tabular}
\end{table} 

\FloatBarrier

\section{Results}
\label{sec:sim_results}

We created synthetic hourly records of $\Theta$, $H_{m0}$ and $T_{m02}$ for $1000$ years with the developed simulation method. We first describe the results for $\Theta$ and then the results for the $H_{m0}$ and $T_{m02}$.

\subsection{Wave Angle Regimes}

%Figure~\ref{fig:wave-direction-sim} shows an example of an observed and a simulated wave direction regime time series. to evaluate the methods ability to generate realistic records of wave angle regimes
To evaluate the method's ability to generate realistic records of $\Theta$, we compare the percentage of time per year in which waves come from the either direction (Southwest or North) in observed and simulated data. We only use $12$ years (2003 - 2014) of the observed record, because the remaining years have gaps, making it more difficult to estimate directional persistence. 

\begin{figure}[thb]
\centering
	\includegraphics[width=0.75\textwidth]{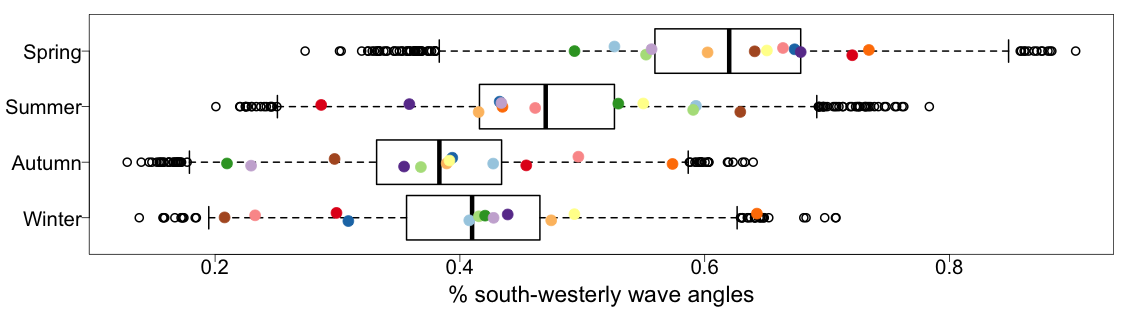}
	\caption{Percentages of time per season with south-westerly wave directions during each of the four seasons. Boxplots summarize data of $10^4$ simulated years. The whiskers represent the lowest and highest simulated data point still within $1.5$ times the interquartile range from the lower and higher quartile, respectively. The open circles represent all remaining data points. Colored points represent observed percentages during $12$ years.}
	\label{fig:wave-directions-boxplot}
\end{figure}

Figure~\ref{fig:wave-directions-boxplot} shows boxplots of percentages of time per season with south-westerly wave directions. The model correctly reflects that the highest percentage of south-western waves occurs in spring, while the lowest percentage of south-western waves occurs in autumn. All except one measured percentage fall within $1.5$ times the interquartile range from the lower and higher quartile, respectively. Also a two-sided, sample-based Kolmogorov-Smirnov (KS) test does not reject the null hypothesis that observed and simulated percentages are equal in distribution for any of the seasons. P-values range from $0.52$ for winter to $0.96$ for autumn. 

%\begin{figure}[thb]
%	\centering 
%	\begin{subfigure}[b]{0.49\textwidth}
%		\includegraphics[width=1\linewidth]{figures/wave_angles/wave_regimes_observed.png}
%		\caption{}
%	\end{subfigure}
%	\begin{subfigure}[b]{0.49\textwidth}
%		\includegraphics[width=1\linewidth]{figures/wave_angles/wave_regimes_simulated.png}
%		\caption{}
%	\end{subfigure}
%	\caption{Wave direction regime time series (a) observed in 2003 and (b) for a simulated year. Light gray represents northern wave directions and dark gray represents south-western wave directions.}
%	\label{fig:wave-direction-sim}
%\end{figure}

\subsection{Significant wave heights and mean zero-crossing periods}
\label{sec:results:wave-angles}

\begin{figure}[thb]
	\centering 
	\begin{subfigure}[b]{0.49\textwidth}
		\includegraphics[width=1\linewidth]{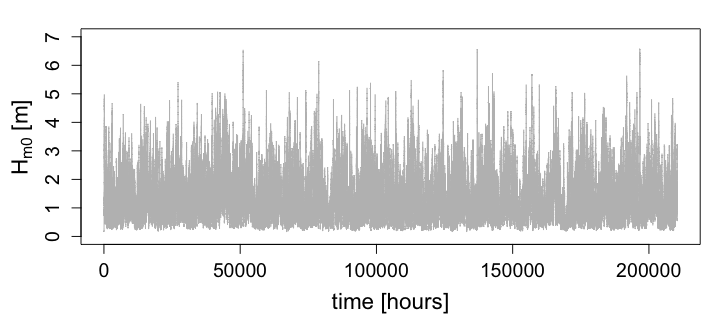}
		\caption{}
			\label{fig:simulated_ts_hs}
	\end{subfigure}
	\begin{subfigure}[b]{0.49\textwidth}
		\includegraphics[width=1\linewidth]{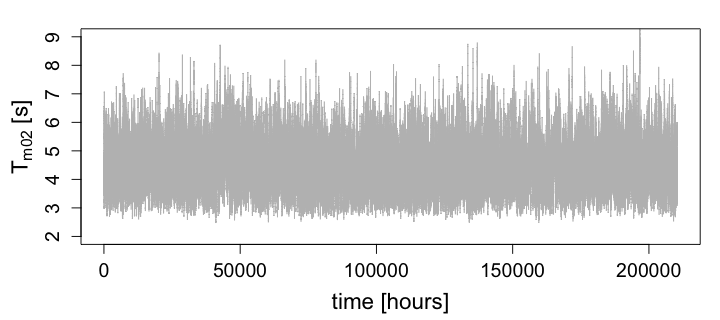}
		\caption{}
			\label{fig:simulated_ts_tp}
	\end{subfigure}
	\caption{Simulated time series of (a) $H_{m0}$ and (b) $T_{m02}$ for a duration of $24$ years.}
\end{figure}

Figures~\ref{fig:simulated_ts_hs} and~\ref{fig:simulated_ts_tp} show examples of the simulated  records for $H_{m_0}$ and $T_{m_{02}}$. The lengths of the simulated series are 24 years, the same as the lengths of the observed series shown in Figures~\ref{fig:ts_x_hs} and~\ref{fig:ts_x_tp}. By visual comparison of the Figures, the simulated series appear to reflect the main characteristics of the observed ones. In particular, the model produces annual cycles and inter-year differences. Nonetheless, the maximum $H_{m0}$ never exceeds its highest observed value, while the maximum mean zero-crossing period does exceeds its highest observed values.

To evaluate the method's ability to generate realistic records of $H_{m0}$ and $T_{m02}$ corresponding the direction regime time series in more depth, we compare the univariate and bivariate empirical densities of simulated hourly values and their persistence above predefined thresholds to the ones observed. For the comparisons, we rely on visual diagnostics instead of statistical goodness of fit tests, because any model would be rejected for a sample size this large: there are $201'960$ observed and $8'766'000$ simulated hourly values.

\subsubsection{Univariate and Bivariate Densities of Hourly Data}

In this section, we present figures of annual densities of the different variables under consideration. In every figure the left panel shows the densities for waves from both directional regimes, the middle panel shows the densities for waves from the Northern regime and the right panel shows the density for waves from the Southwestern regime.

\begin{figure}[thb]
 	\centering 
   	\begin{subfigure}[b]{0.32\textwidth}
   		\includegraphics[width=1\linewidth]{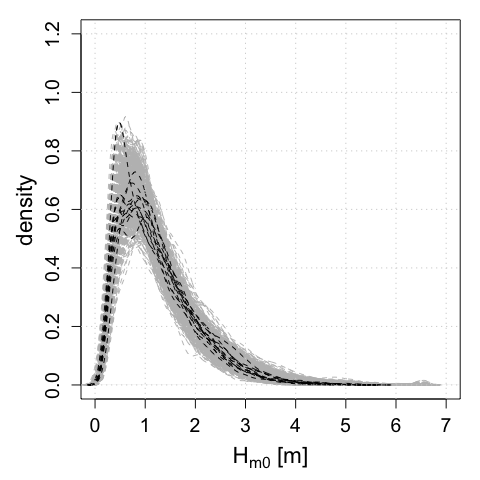}
   		\caption{}
   		\label{fig:hs_yearly_density_all} 
   	\end{subfigure}
    	\begin{subfigure}[b]{0.32\textwidth}
    		\includegraphics[width=1\linewidth]{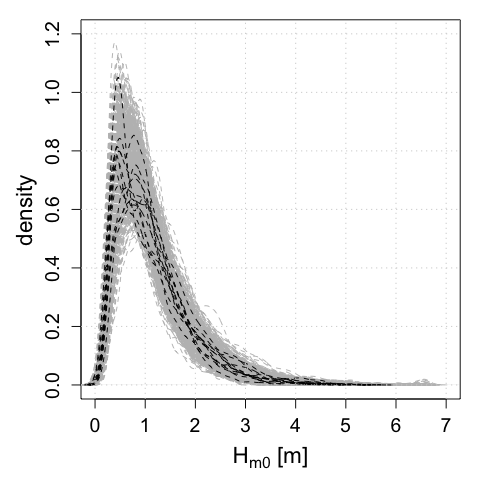}
    	 	\caption{}
    	   	\label{fig:hs_yearly_density_N} 
    	\end{subfigure}
   	  	\begin{subfigure}[b]{0.32\textwidth}
   	  		\includegraphics[width=1\linewidth]{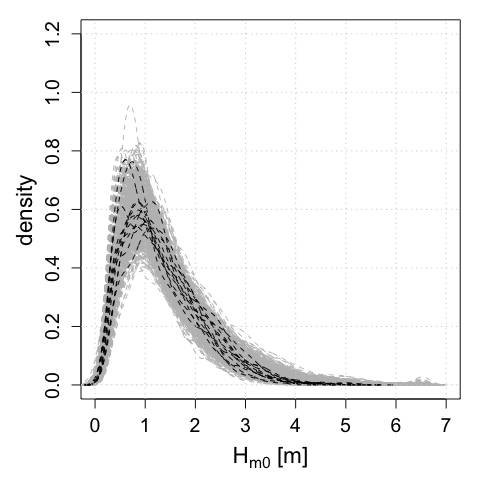}
   	  	   	\caption{}
   	  	   	\label{fig:hs_yearly_density_SW} 
   	  	\end{subfigure}
   	\caption{Univariate densities of $H_{m0}$ for waves (a) from either directional regime, (b) from the Northern regime and (c) from the Southwestern regime.}
   	\label{fig:hs_yearly_density}
\end{figure}

Figure~\ref{fig:hs_yearly_density} shows the annual empirical probability densities of  hourly $H_{m0}$ for 1000 simulated years together with the annual empirical probability densities the 24 observed years. The simulation model can reproduce the characteristics of the densities of the observed data. The densities of the simulated series form a cloud around the densities of the observed series.  The differences between annual densities are notable in all directions in simulated and observed series. Moreover, northern waves tend to be higher than southwestern waves and have a more narrow distribution in both cases. 

\begin{figure}
   	\begin{subfigure}[b]{0.32\textwidth}
   		\includegraphics[width=1\linewidth]{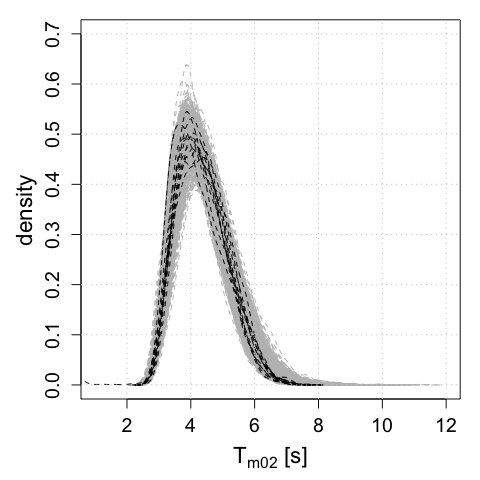}
   		\caption{}
   		\label{fig:tp_yearly_density_all} 
   	\end{subfigure}
    	\begin{subfigure}[b]{0.32\textwidth}
    		\includegraphics[width=1\linewidth]{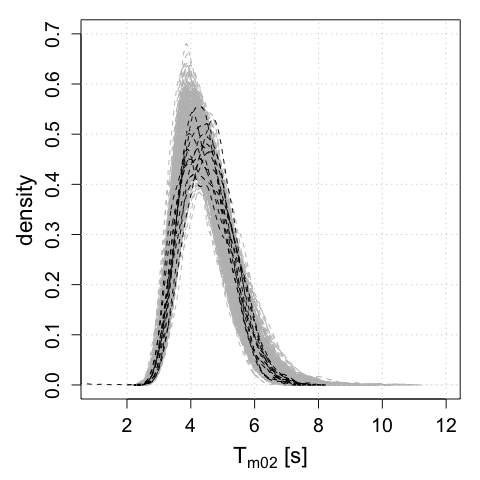}
    		\caption{}
    		\label{fig:tp_yearly_density_N} 
    	\end{subfigure}
   	  	\begin{subfigure}[b]{0.32\textwidth}
   	  		\includegraphics[width=1\linewidth]{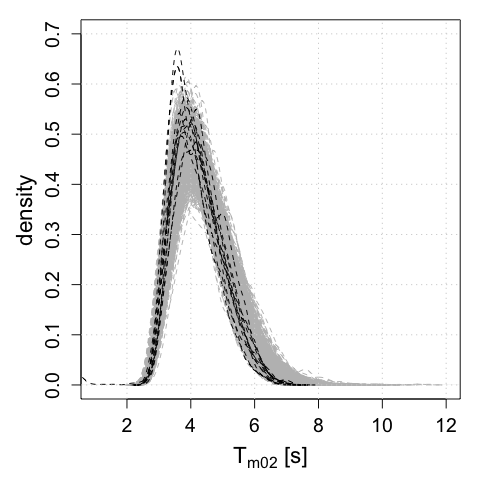}
   	  		\caption{}
   	  		\label{fig:tp_yearly_density_SW} 
   	  	\end{subfigure}
   	  	\caption{Univariate densities of $T_{m02}$ for waves (a) from either directional regime, (b) from the Northern regime and (c) from the Southwestern regime.}
   	  	\label{fig:tp_yearly_density}
   	\end{figure}

Similarly, Figure~\ref{fig:tp_yearly_density} shows the densities for $T_{m02}$. Again, annual differences are notable in the observed series, but differences between directional regimes are less pronounced. This is captured by the simulated series for all, but two years. In these cases, the densities of observed values do not fall within the cloud of densities of simulated values. This is most pronounced for the mode of the density for waves from the southwestern direction.

\begin{figure}
   	\begin{subfigure}[b]{0.32\textwidth}
   		\includegraphics[width=1\linewidth]{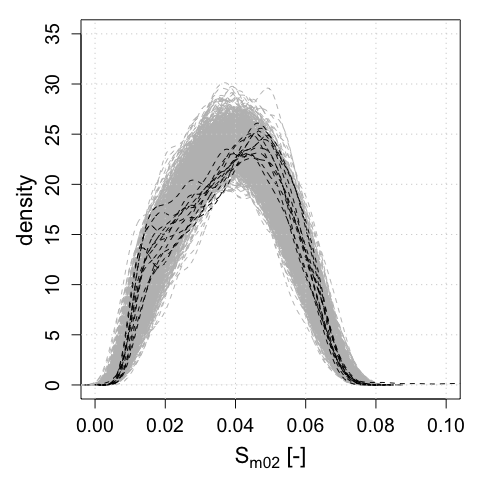}
   		\caption{}
   		\label{fig:sp_yearly_density_all} 
   	\end{subfigure}
    	\begin{subfigure}[b]{0.32\textwidth}
    		\includegraphics[width=1\linewidth]{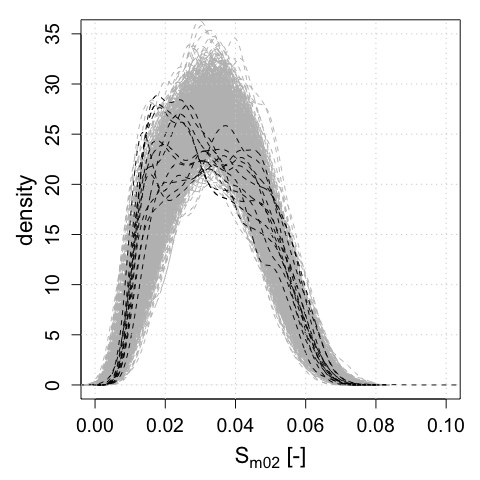}
    		\caption{}
    		\label{fig:sp_yearly_density_N} 
    	\end{subfigure}
   	  	\begin{subfigure}[b]{0.32\textwidth}
   	  		\includegraphics[width=1\linewidth]{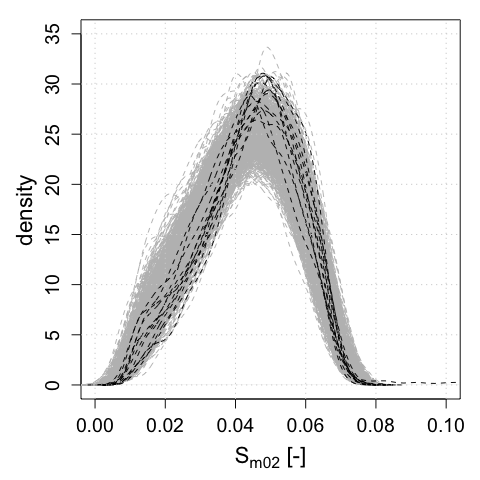}
   	  		\caption{}
   	  		\label{fig:sp_yearly_density_SW} 
   	  	\end{subfigure}
   	   	  	\caption{Univariate densities of steepness for waves (a) from either directional regime, (b) from the Northern regime and (c) from the Southwestern regime.}
   	   	  	\label{fig:sp_yearly_density}
\end{figure}

Next, we computed the steepness of waves. This is another important property of waves, which depends on both $H_{m0}$ and $T_{m02}$ (equation~\ref{eq:wave_steepness}). Figure~\ref{fig:sp_yearly_density} shows the densities for wave steepness. While the densities of the simulated values form a cloud around almost all densities of the observed values, they appear to have different characteristics by visual comparison. Notable is that some of the densities of the observed values for the northern regime appear to have bimodal densities. This is not the case for densities of simulated values (To see this, we inspected them one by one).

\begin{figure}[thb]
  	\centering  	
    \begin{subfigure}[b]{0.32\textwidth}
    	    \includegraphics[width=1\linewidth]{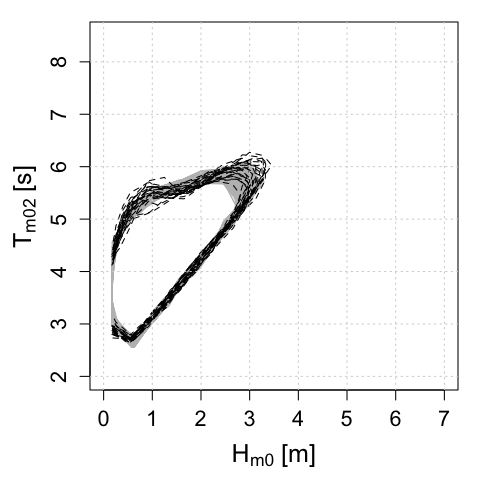}
    	    \caption{}
    	    \label{fig:2d_yearly_05density_all} 
    \end{subfigure}
    \begin{subfigure}[b]{0.32\textwidth}
    	     \includegraphics[width=1\linewidth]{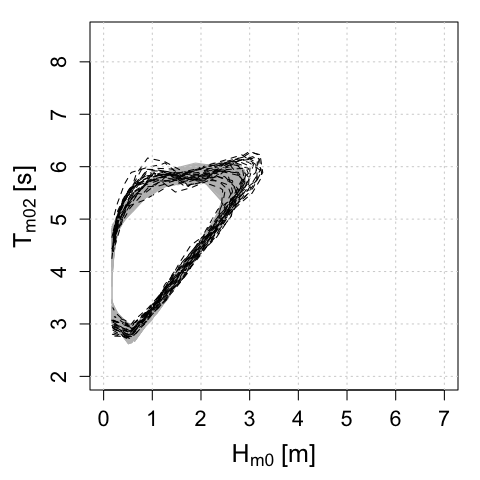}
    	     \caption{}
    	     \label{fig:2d_yearly_05density_N} 
    \end{subfigure}
    \begin{subfigure}[b]{0.32\textwidth}
    	   	\includegraphics[width=1\linewidth]{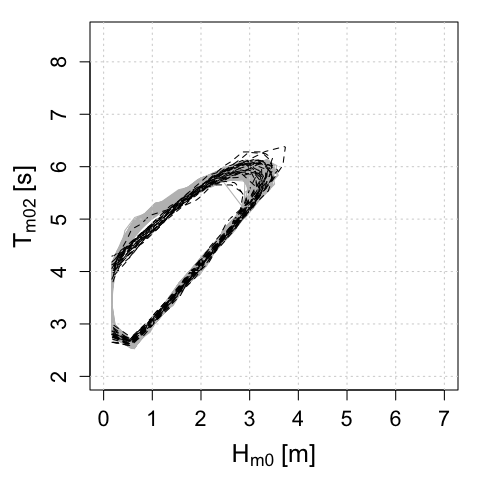}
    	   	\caption{}
    	   	\label{fig:2d_yearly_05density_SW} 
    \end{subfigure}
	\caption{Bivariate annual $5 \cdot 10^{-2}$ density contour of $H_{m0}$ and $T_{m02}$ for waves (a) from either directional regime, (b) from the Northern regime and (c) from the Southwestern regime.}
	\label{fig:2d_yearly_05density} 
\end{figure}

\begin{figure}
     \begin{subfigure}[b]{0.32\textwidth}
     	\includegraphics[width=1\linewidth]{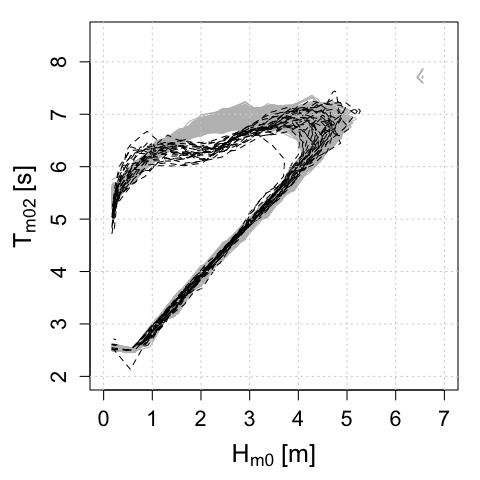}
     	\caption{}
     	\label{fig:2d_yearly_005density_all} 
     \end{subfigure}
      \begin{subfigure}[b]{0.32\textwidth}
     		\includegraphics[width=1\linewidth]{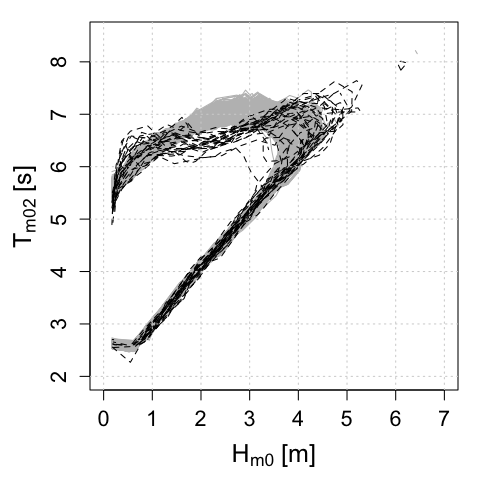}
	 		\caption{}
     		\label{fig:2d_yearly_005density_N} 
      \end{subfigure}
      \begin{subfigure}[b]{0.32\textwidth}
       		\includegraphics[width=1\linewidth]{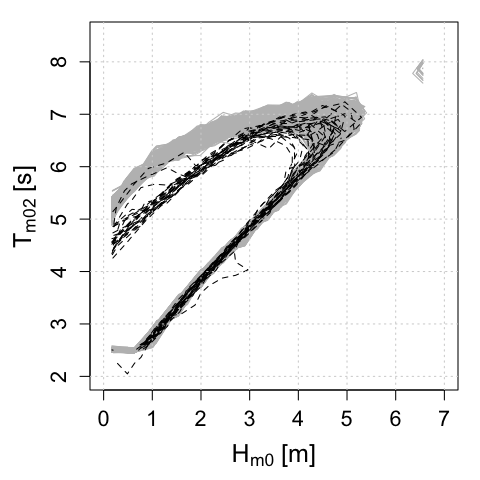}
       		\caption{}
       		\label{fig:2d_yearly_005density_SW} 
     \end{subfigure}
	\caption{Bivariate annual $5 \cdot 10^{-3}$ density contour of $H_{m0}$ and $T_{m02}$ for waves (a) from either directional regime, (b) from the Northern regime and (c) from the Southwestern regime.}
	\label{fig:2d_yearly_005density} 
 \end{figure}

Finally, we compared bivariate densities of hourly values. Figure~\ref{fig:2d_yearly_05density} shows annual contour lines for values with density $5 \cdot 10^{-2}$ and  Figure~\ref{fig:2d_yearly_005density} shows annual contour lines for values with density $5 \cdot 10^{-3}$.

The model appears to approximate the $5 \cdot 10^{-2}$ density contours well. In particular, the maximum steepness condition is well-represented. Nonetheless, a couple of features are not accurately represented. On one hand, the joint maxima of $H_{m0}$ and $T_{m02}$ at this density level tend to be underestimated by the model. On the hand, 
$T_{m02}$ tends to be higher than observed for $H_{m0} < 2m$ for waves from the southwestern regime.

The model approximates the $5 \cdot 10^{-3}$ density contours less well than the $5 \cdot 10^{-2}$ density contours. When we are not distinguishing between the directional regimes, the model still produces realistic contours, though $T_{m02}$ tends to be overestimated for values $H_{m0}$ around $2m$. For wave from the northern regime, this is the case as well. In addition, the joint maxima of $H_{m0}$ and $T_{m02}$ at this density level end to be underestimated. For waves from the southwestern regime, $T_{m02}$ tends to be overestimated for values of $H_{m0}$ smaller than $3m$. On the other hand, joint maxima of $T_{m02}$ and $H_{m0}$ are capture well.

%In general, 
%
%Table~\ref{tab:spearmans_rho}
%
%An overall tendency to underestimate the degree of positive dependence for southwestern directions is also reflected by values of Spearman's $\rho$ (see ). 
%
%\begin{table}[ht]
% \caption{Values of Spearman's $\rho$ for observed and simulated $H_{m0}$ and $T_{m02}$ for different wave angle directions.}
% \label{tab:spearmans_rho}
% \centering
% \begin{tabular}{c c c }
% \toprule
% Wave direction & Observed data & Simulated data \\ 
% \midrule
% All & $0.74$ & $0.73$ \\
% Northern & $0.70$ & $0.70$ \\
% Southwestern & $0.88$ & $0.80$ \\
% \bottomrule
% \end{tabular}
%\end{table}

\subsubsection{Storm durations and interarrival times}

In this section, we investigate the capability of the model to simulate sequences of storms. We focus on storm durations and interarrival times. In the literature these are often referred to as persistence regimes above and below predefined thresholds. We analyzed persistence for six pairs of thresholds for $H_{m0}$ and $T_{m02}$: both variables jointly exceeding their respective univariate $0.8$, $0.9$, $0.95$, $0.965$, $0.975$, and $0.99$ quantiles. The corresponding values are listed in Table~\ref{tab:storm_definitions}

 \begin{table}[ht]
	\caption{Quantiles of $H_{m0}$ and $T_{m02}$ that are selected as critical threshold values for the analysis of storm durations and interarrival times.}
	\label{tab:storm_definitions}
	\centering
	\begin{tabular}{c c c c c c c}
		\toprule
		Quantile & $0.8$ & $0.9$ & $0.95$ & $0.965$ & $0.975$ & $0.99$ \\ \midrule
		$H_{m0}$ [m] & $1.83$ & $2.34$ & $2.78$ & $3$ & $3.19$ & $3.70$ \\
		$T_{m02}$ [s] & $5.1$ & $5.5$ & $5.8$ & $6.0$ & $6.1$ & $6.4$\\ 
		\bottomrule
	\end{tabular}
\end{table}

The different quantiles were chosen so that they would represent a wide range of conditions. The choice of the $0.965$ quantile is motivated by\cite{li2014probabilistic} who simulated sea storms for a location in the Dutch North sea with comparable geographical properties. They use a $H_{m0}$ threshold of $3$m in combination with a surge threshold of $0.5$m. They chose these thresholds, because more severe conditions are likely to cause morphological change \citep{Quartel2007}. This value of $H_{m0}$ corresponds to the $96.5\%$ quantile in our data set, which is why we selected it as well.

\begin{figure}[t]
	\centering 
	\begin{subfigure}[t]{0.32\textwidth}
		\includegraphics[width=1\linewidth]{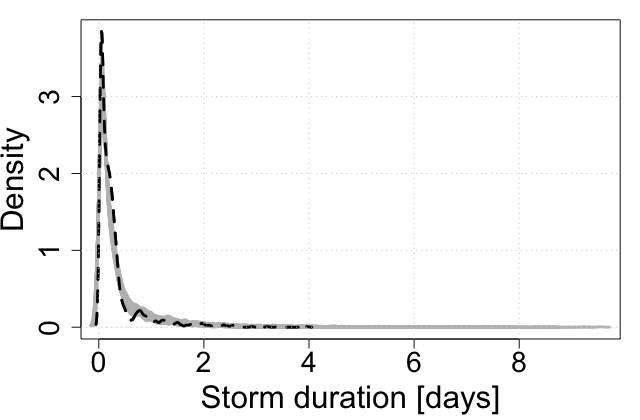}
		\caption{$H_{m0} \geq 1.83$m and $T_{m02} \geq 5.1$s.}
	\end{subfigure} 
	\begin{subfigure}[t]{0.32\textwidth}
		\includegraphics[width=1\linewidth]{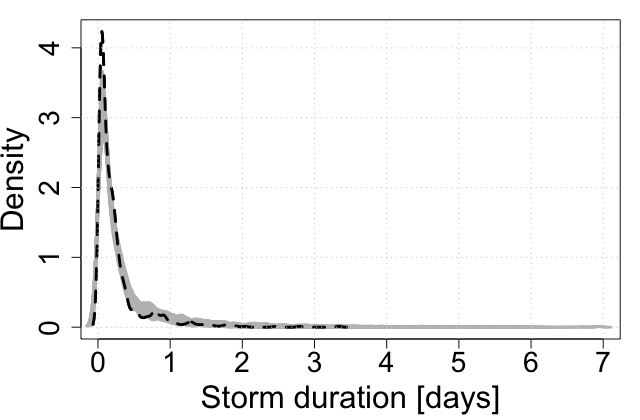}
		\caption{$H_{m0} \geq 2.34$m and $T_{m02} \geq 5.5$s.}
	\end{subfigure} 
	\begin{subfigure}[t]{0.32\textwidth}
		\includegraphics[width=1\linewidth]{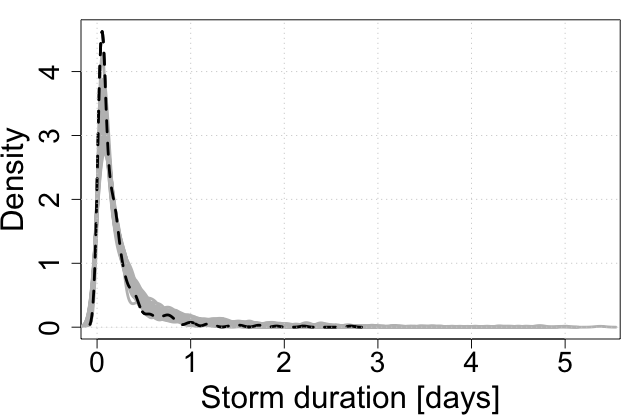}
		\caption{$H_{m0} \geq 2.78$m and $T_{m02} \geq 5.8$s.}
	\end{subfigure} 

	\begin{subfigure}[t]{0.32\textwidth}
		\includegraphics[width=1\linewidth]{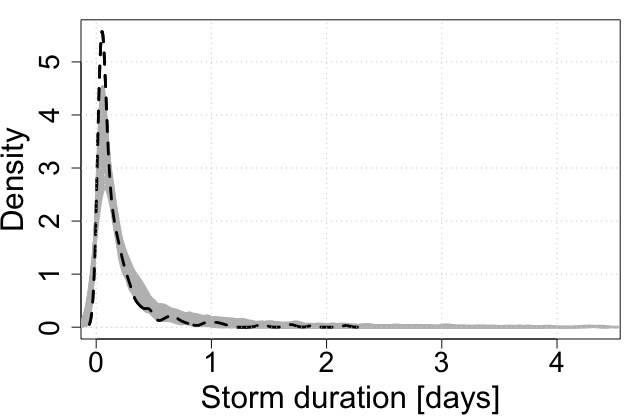}
		\caption{$H_{m0} \geq 3.00$m and $T_{m02} \geq 6.0$s.}
	\end{subfigure} 
	\begin{subfigure}[t]{0.32\textwidth}
		\includegraphics[width=1\linewidth]{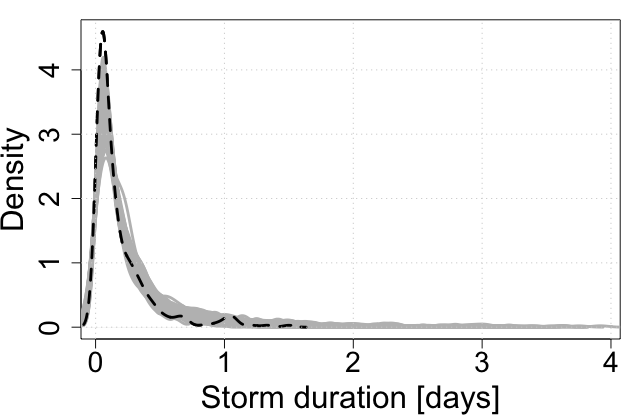}
		\caption{$H_{m0} \geq 3.19$m and $T_{m02} \geq 6.1$s.}
	\end{subfigure} 
	\begin{subfigure}[t]{0.32\textwidth}
		\includegraphics[width=1\linewidth]{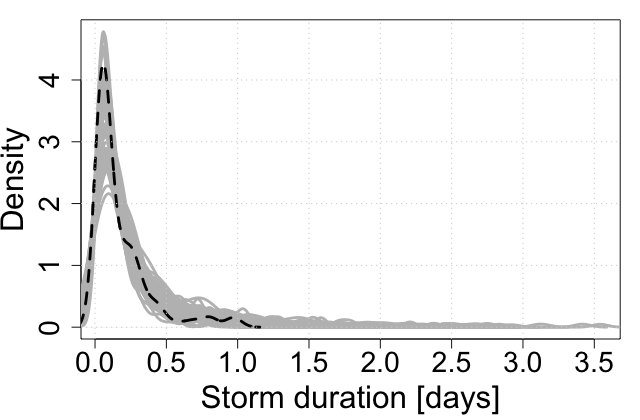}
		\caption{$H_{m0} \geq 3.70$m and $T_{m02} \geq 6.4$s.}
	\end{subfigure} 
	\caption{Persistence of $H_{m0}$ and $T_{m02}$ above different thresholds.}
	\label{fig:storm_durations}
\end{figure}

\begin{figure}[t]
	\centering 
	\begin{subfigure}[t]{0.32\textwidth}
		\includegraphics[width=1\linewidth]{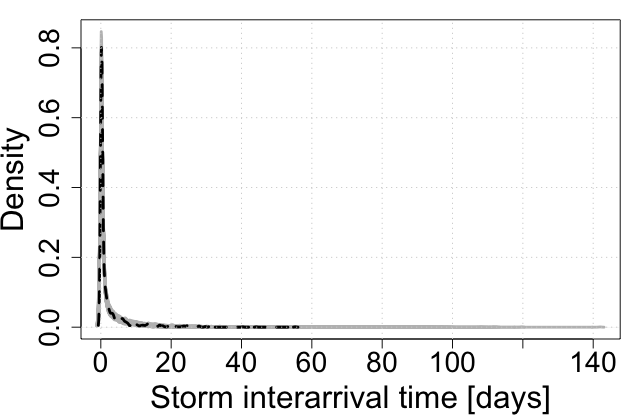}
		\caption{$H_{m0} \geq 1.83$m and $T_{m02} \geq 5.1$s.}
	\end{subfigure} 
	\begin{subfigure}[t]{0.32\textwidth}
		\includegraphics[width=1\linewidth]{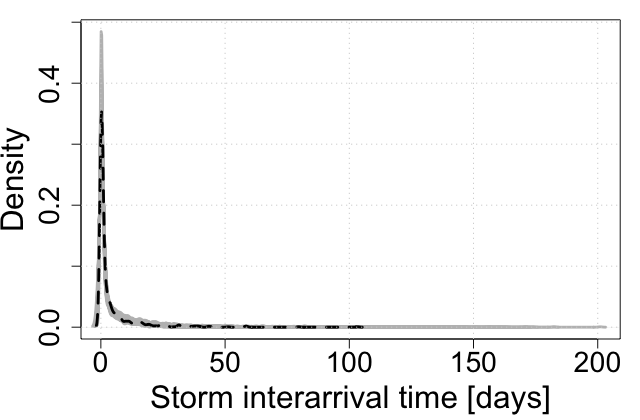}
		\caption{$H_{m0} \geq 2.34$m and $T_{m02} \geq 5.5$s.}
	\end{subfigure} 
	\begin{subfigure}[t]{0.32\textwidth}
		\includegraphics[width=1\linewidth]{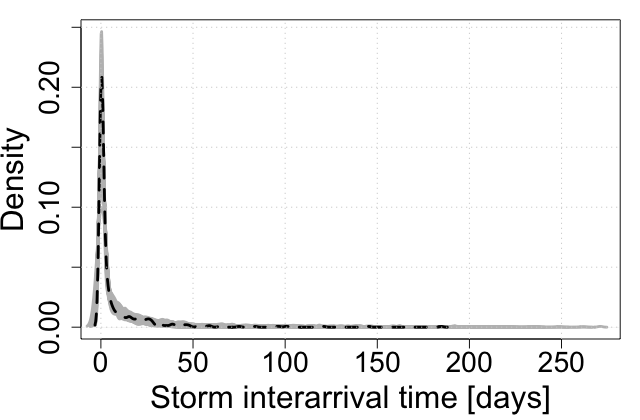}
		\caption{$H_{m0} \geq 2.78$m and $T_{m02} \geq 5.8$s.}
	\end{subfigure} 
	
	\begin{subfigure}[t]{0.32\textwidth}
		\includegraphics[width=1\linewidth]{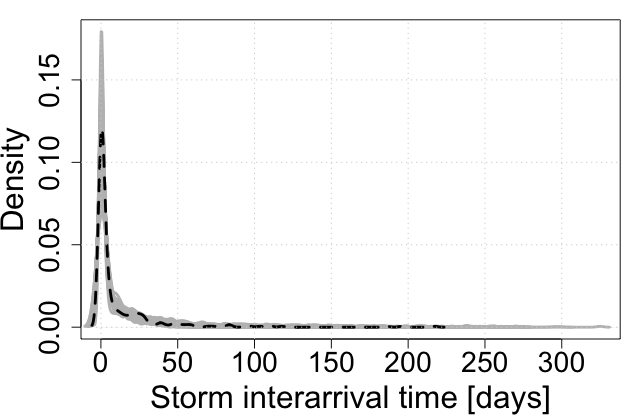}
		\caption{$H_{m0} \geq 3.00$m and $T_{m02} \geq 6.0$s.}
	\end{subfigure} 
	\begin{subfigure}[t]{0.32\textwidth}
		\includegraphics[width=1\linewidth]{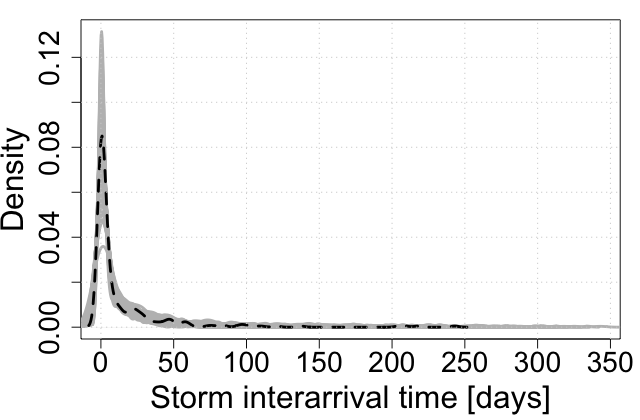}
		\caption{$H_{m0} \geq 3.19$m and $T_{m02} \geq 6.1$s.}
	\end{subfigure} 
	\begin{subfigure}[t]{0.32\textwidth}
		\includegraphics[width=1\linewidth]{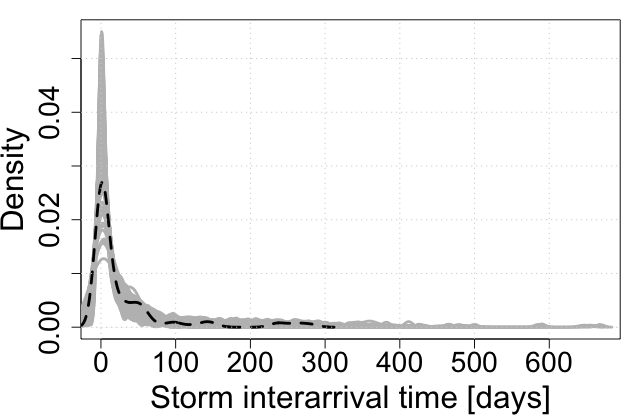}
		\caption{$H_{m0} \geq 3.70$m and $T_{m02} \geq 6.4$s.}
	\end{subfigure} 
	\caption{Persistence of $H_{m0}$ and $T_{m02}$ below different thresholds.}
	\label{fig:storm_ias}
\end{figure}

Figure~\ref{fig:storm_durations} shows densities of storm durations and Figure~\ref{fig:storm_ias} shows densities of storm interarrival times in observed and simulated data. The six panels correspond to the six pairs of thresholds listed in Table~\ref{tab:storm_definitions}. Each panel shows a density that has been computed from a 12-year segment of the observed time series (the same one that has been analyzed in Section~\ref{sec:results:wave-angles}) and 83 densities that have been computed from separate $12$-year long segments of the $1000$-year long simulated time series.

In general, the simulation model produces realistic storm durations and interarrival times for all pairs of thresholds. In the case of storm durations, the density computed from observed values lies within the cloud of densities computed from simulated values for the lowest and the highest pair of thresholds. For the other pairs, the mode of the observed density is higher than any mode of the simulated densities. Furthermore, some simulated storm durations are approximately two to three times as long as the longest observed storm durations, depending on the pair of thresholds. In the case of the storm interarrival times, the densities computed from observed values lie within the cloud of densities computed from simulated values for all pairs of thresholds. Similarly to the storm durations, some of simulated storm interarrival times are approximately one and a half to three times as long as the longest observed storm interarrival times, depending on the pair of thresholds.

Furthermore, the model simulates a realistic number of storms, when comparing the observed $12$ year segment to the $83$ simulated $12$ year segments. Of course, this is expected given the results on storm duration and interarrival time. For all pairs of thresholds the observed number of storms lies within the $5\%$- and the $95\%$- quantile of simulated number of storms. The exact values are reported in Table~\ref{tab:number_of_storms}.

\begin{table}[ht]
	\caption{Observed number of storms between 2003 and 2014, and $5\%$- and $95\%$- quantiles of simulated number of storms for different thresholds.}
	\label{tab:number_of_storms}
	\centering
	\begin{tabular}{c c c c c c c}
		\toprule
		Threshold-defining quantile & $0.8$ & $0.9$ & $0.95$ & $0.965$ & $0.975$ & $0.99$ \\ \midrule
		Observed number of storms & $3307$ & $1937$ & $1151$ & $805$ & $609$ & $279$ \\ 
		$5\%$  quantile of simulated number of storms & $2832$ & $1543.4$ & $887.6$ & $613.8$ & $473.2$ & $196.8$ \\
		$95\%$  quantile of simulated number of storms & $3338$ & $1946.6$ & $1173.9$ & $874.8$ & $694.6$ & $330.6$\\ 
		\bottomrule
	\end{tabular}
\end{table}

%\begin{figure}[thb]
%	\centering 
%	\begin{subfigure}[b]{1\textwidth}
%		\includegraphics[width=1\linewidth]{figures/storm_durations.png}
%		\caption{}
%	\end{subfigure}
%	
%	\begin{subfigure}[b]{1\textwidth}
%		\includegraphics[width=1\linewidth]{figures/storm_interarrivals.png}
%		\caption{}
%	\end{subfigure}
%	\caption{Boxplots of storm durations and storm interarrival times for storm I and storm II events. The underlying observed and simulated time series from which the durations and interarrival times were derived were both $12$ years long.}
%	\label{fig:storm_durations}
%\end{figure}

%As mentioned in the introduction, a potential advantage of simulating full time series is that no prior assumptions have to be made on critical threshold values or storm shape. This is useful, because the assumptions vary per application. 

\FloatBarrier
\section{Discussion}
\label{sec:discussion}

While the simulation method is suitable to generate time series that exhibit statistical features relevant for coastal and offshore risk analysis, it still has limitations that affect its applicability. We discuss the main limitations in this section. 

The modeling of the wave direction has been simplified by assuming a categorical variable with two states which representing two directional sectors: north and southwest. The main advantage of this approach is that difficulties related to the circular nature of the variable can be avoided. Moreover, the results in Section~\ref{sec:results:wave-angles} show that time series of $\Theta$ can be modeled accurately as a seasonal renewal process. However, this simplification can be a limitation for practical applications, because the wave direction affects structural loading as well as sediment transport: Loads are highest for waves that hit the structure under normal angles (e.g., \cite{Goda2008}) whereas erosion can be significantly higher for waves that hit the coast under non-normal than normal angles (e.g., \cite{Heijer2013}). Hence, it would be desirable to further develop the proposed methodology in order to simulate waves at a higher directional resolution. Eventually, additional variables, such as wind speeds and surges should be included as well to broaden the applicability of the method. Wind speeds are often crucial for offshore operations (e.g., \cite{leontaris2016probabilistic}), while surges are central to coastal risk assessments and the prediction of longterm morphological changes (e.g. \cite{li2014probabilisticestimation,davies2017Improved}).

The choice of using two regimes for the wave direction was inspired by its clear bimodal distribution (cf. Figure~\ref{fig:histogram-WA}) and the univariate and joint densities of hourly values of $H_{m0}$ and $T_{m02}$ being different in the two modes. We tried to capture the difference by using a regime-switching joint distribution for the residuals of the ARMA models corresponding to $H_{m0}$ and $T_{m02}$. With this approach, we were able to capture part of the difference, but not all. In particular, the approach worked well for the univariate densities of hourly values of $H_{m0}$. The univariate densities of hourly values of $T_{m02}$ are also captured, but the difference across regimes is less noticeable for this variable. Nonetheless, the bi-modality of the densities of hourly values of $S_{m0}$ in the northern regime could not be represented. In terms of bivariate density, the model produced differences between the regimes, but they were not as pronounced as the ones observed. Hence, it does not appear to be sufficient to rely on regime switches in the joint residual distribution to capture the difference between northern and southwestern wave conditions. To improve this, future research could explore regime-switching ARMA parameters $p$ and $q$ and the extension to a vector-ARMA. However, investigating these options would require the derivation and implementation of parameter estimation procedures. As far as we know, estimation procedures have only been implemented for AR processes, but not for ARMA processes (e.g., \cite{tsdyn2018,mswm2018}).

Another limitation of the methodology is due to the use of the empirical cumulative distribution function in the initial normalization of the data (cf. equation \ref{eq:PIT}): In simulations values of $H_{m0}$ are obtained by applying an inverse PIT to simulated values of $Z^{(H_{m0})}$ using the inverse empirical cumulative distribution function of $H_{m0}$. Hence, the simulated values of $H_{m0}$  span the same range as observed values of $H_{m0}$. In other words, we will never simulate more extreme values of $H_{m0}$ than we have observed. The case of $T_{m02}$ is different, because simulated values of $\widetilde{T}_{m02}$, and not of $T_{m02}$, are obtained from simulated values of $Z^{(T_{m02})}$ via the inverse PIT. Hence, simulated values of $\widetilde{T}_{m02}$ span the same range as the corresponding observed values. However, next \begin{equation}
{T}_{m02} = {T}_{m02_{min}} + \widetilde{T}_{m02}
\end{equation} (cf. equation \ref{eq:TP_trafo}) and higher values than observed can arise for ${T}_{m02}$ for certain combinations of ${T}_{m02_{min}}$ and $\widetilde{T}_{m02}$. This behavior in the extremes can be recognized in the simulated time series shown in Figures~\ref{fig:simulated_ts_hs} and~\ref{fig:simulated_ts_tp}. Since we were aware of this limitation, we did not investigate the model behavior for extreme values, such as persistence above quantiles larger than $0.99$ or bivariate contours with density lower than $5\cdot 10^{-3}$. Nonetheless, finding an alternative variable transformation that overcomes this limitation and exploring the methods skill to simulate extremes is relevant for applications related to the design of infrastructures or reliability analyses.

\FloatBarrier
\section{Conclusion}
\label{sec:conclusion}

In this paper, we presented a simulation method for joint time series of $H_{m0}$, $T_{m02}$ and $\Theta$. The latter is a categorical variable that distinguishes northern and southwestern waves. Time series can be simulated at a high resolution of $1$ hour, which is useful for risk analyses in various coastal and offshore applications. The method has been applied to a data set in the Dutch North sea.

The method contains several modeling steps and relies on renewal processes, Fourier series with random coefficients, ARMA processes, copulas, and regime-switching. A particular feature is a data-driven estimate for a wave height-dependent limiting wave-steepness condition, which we use to describe part of the dependence between $H_{m0}$ and $T_{m02}$ and which facilitates the copula-based dependence modeling later on. Similar to many other studies, annual seasonality is represented by Fourier series. The coefficients are modeled as inter-dependent random variables to account for inter-year differences. At this point we did not consider climatic covariates and recommend to examine, if they have predictive skill for inter-year differences. 

The stationary components of the two processes are represented as ARMA with a regime-switching joint residual distribution, constructed with copulas. The regime-switches are triggered by switches in wave direction from North to Southwest. While these regime-switches result in differences between bivariate distributions of simulated $H_{m0}$ and $T_{m02}$ when conditioned on the northern and southwestern regime, they are not as pronounced as in observed data. We recommend that future research is directed at improving and extending the simulation method as to better capture these differences. Nonetheless, the unconditioned bivariate distribution of $H_{m0}$ and $T_{m02}$ appears to be represented adequately. 

Moreover, storm durations and storm interarrival times are well captured for two different storm definitions, which rest on different critical threshold values for $H_{m0}$ and $T_{m02}$. As storm sequences are adequately represented, the model has potential value for applications in coastal and offshore engineering, such as the prediction of long-term morphological changes, or the planning and budgeting of offshore operations.

\FloatBarrier
\appendix

\section{Autoregressive moving-average (ARMA) models}
\label{sec:background-arma}

ARMA models provide a parsimonious description of weakly stationary time series. For a comprehensive introduction to the topic see, for example,  \cite{Box2015},  \cite{Brockwell2016} or \cite{Shumway2017}.  A stochastic process $\lbrace Z_t: t=1,2,3,...\rbrace$ is considered to be weakly stationary if all its moments up to the order of two do not vary in time. Thus, the mean and the variance of random variable $Z_t$ is equal to a constant and the covariance between any pair $\lbrace Z_t, Z_{t+k} \rbrace,  \forall k \in \mathbb{N}$, only depends on $k$ but not on $t$.  

A process $Z_t$ is called ARMA, if it can be expressed as the following function of past observations, $Z_{t-1}, .., Z_{t-p}$, and past residuals, $\epsilon_{t-1}, .., \epsilon_{t-q}$: \begin{equation}
Z_t = c + \sum_{j=1}^{p} \phi_j Z_{t-j} + \epsilon_t + \sum_{j=1}^{q} \theta_j \epsilon_{t-j},
\end{equation} where $c$ is a constant intercept term, $\phi_j$ and $\theta_j$ are non-zero constants, and the residuals $\epsilon_t$ are independent and identically distributed (i.i.d.) with zero mean. If every $\phi_j$ is zero, the process is said to be a moving average process of order $q$, $MA(q)$, and if every $\theta_j$ is zero, then it is called an autoregressive process of order $p$,  $AR(p)$. 

For given orders $p$ and $q$, the model parameters, $\phi_j$ and $\theta_j$, can be estimated by maximum likelihood  or by minimizing the conditional sum of squares of the fitted residuals. An indication for suitable orders can usually be found by inspecting the autocorrelation function (ACF) and the partial ACF (PACF). The ACF at lag $k$ is defined as \begin{equation}
\rho(k) = corr(X_{t+k}, X_t), 
\end{equation} where $corr$ denotes the product moment correlation. In contrast, the PACF measures the correlation between $X_{t+k}$ and $X_t$, for $k \geq2$, with the linear effects of $X_{t+1}, ..., X_{t+k-1}$ removed. In order to define the PACF, let $\hat{X}_{t+k}$ denote the estimated mean from a regression of $X_{t+k}$ on $\lbrace X_{t+k-1}, ..., X_{t+1} \rbrace$ and $\hat{X}_{t}$ denote the estimated mean from a regression of $X_{t}$ on $\lbrace X_{t+1}, ..., X_{t+k-1} \rbrace$. The PACF for lag $k$ can then be defined as: \begin{equation}
\phi_{kk} = 
\begin{cases}
corr(X_1,X_0), \quad k=1 \\
corr(X_{t+k} -\hat{X}_{t+k},X_{t} -\hat{X}_{t} ), \quad k \geq 2
\end{cases}.
\end{equation}

ARMA models with different orders have distinctive ACF and PACF behaviors. The ACF of an $AR(p)$ process decays slowly, while its PACF has a cut off at lag $p$. Conversely, the ACF of an $MA(q)$ process has a cut off at lag $q$, but its PACF decays more slowly. Finally, both ACF and PACF tail off in $ARMA(p,q)$ processes and are dominated by mixtures of exponentials and damped sine waves after the first $q-p$ lags and $p-q$ lags respectively.

\section{Copulas}
\label{sec:background-copulas}

Copula models are used to represent the joint behavior of several random variables. With a copula approach, the main limitation that is encountered with classical families of bivariate distributions (e.g., Gaussian, student-t, Gamma or generalized extreme value) is avoided. The limitation is that the individual behavior of the of variables must be characterized by the same family of univariate distributions as the joint distribution. For example, if two variables are joint normally distributed, each of them must follow a univariate normal distribution as well. However, in many practical applications the joint distribution of variables which follow different univariate distributions is sought.

A copula describes the dependence between random variables, separately from their respective marginal behaviors. The underlying theory is based on Sklar's theorem (\cite{Sklar1959fonctions}), which states a joint distribution function, $F$, of random variables, $X_1, ..., X_d$, with univariate distribution functions, $F_1,..., F_d$, can be represented by a copula, $C$, in this way:
 \begin{equation}
F(\mathbf{x}) = C(F_1(x_1),..., F_d(x_d)), \qquad\mathbf{x} \in \mathbb{R}^d.
\end{equation} The copula itself is a $d$-variate distribution function on $[0,1]^d$ with uniform margins. Thus, a valid model for $F$ can be constructed from appropriate models for $F_1, ..., F_d$ and for $C$.

In this article we focus on the bivariate case. In the literature, many parametric copula families have been proposed, covering a wide range of dependence structures including tail dependences. For instance, \cite{Nelsen2006, joe2014dependence, Durante2015} provide a comprehensive theoretical overview, while, for example, \cite{genest2007everything,Salvadori2007} provide a good introduction for engineering purposes. In particular, guidelines for using copulas in maritime engineering are illustrated in \cite{salvadori2014practical,salvadori2015practical}.

\section{Statistical limiting wave steepness condition}
\label{app:wave-steepness}

Visual inspection of the data indicated that the limiting wave steepness is not constant, but varies with $H_{m0}$, as described in the main text in Section~\ref{sec:method-steepness}. To account for this behavior we fit the curve \begin{equation}
s_{{m02}_{max}}(H_{m0}) = a \left( \frac{H_{m0}}{b} \right) ^{\left( \frac{c}{H_{m0}} \right)}, \quad a,b,c >0
\end{equation} to the data. 

The scatter plot in Figure~\ref{fig:scatter_hs_sp} shows a horizontal asymptote roughly below $s_{{m02}_{max}}=0.08$, while the observed $s_{{m02}_{max}}$ is rapidly decreasing for small $H_{m0}$. This motivates the functional form of the curve: Suppose $b$ is a large value that cannot be attained by measurements of $H_{m0}$ at the Europlatform. Then, $a$ can be interpreted as the value defining the horizontal asymptote, since $s_{{m02}_{max}} \rightarrow a$, as $H_{m0} \rightarrow b$. Finally, $c$ affects the slope of $s_{{m02}_{max}}$ for smaller values of $H_{m0}$.

The procedure to fit the $s_{{m02}_{max}}$-curve was the following: First, we discretized the $H_{m0}$ data into $108$ discrete bins. These were not equally spaced, but contained an equal number of data points ($1870$). Most bins cover a range in height of $1cm$ or $2cm$. An exception is the widest bin, which spans from $274$cm to $656$cm. Next, we computed the maximum value of 
$s_{{m02}_{max}}$ in each bin and associated it with the value of $H_{m0}$ at the bin center. These data points are shown as orange circles in Figure~\ref{fig:scatter_hs_sp}. Finally, we estimated the coefficients $a$, $b$ and $c$ using nonlinear least-squares. The resulting estimates are $a=0.0782$, $b=999.4$cm (the upper bound was fixed at $1000$) and $c=7.674$cm. The coefficient of determination is $R^2_{adj} = 0.858$ and the root mean square error is $RMSE = 0.0051$. 

According to this limiting steepness condition, $160$ measurements are classified as anomalies, because they are too steep. These are $67$ more than initially identified by visual inspection and amount to less than $0.08\%$ of the data. These data were classified as anomalies and substituted with missing values.

\section*{Acknowledgements}
This work was supported by the European Community's 7th Framework Programme through the grant to RISC-KIT ("Resilience-increasing Strategies for Coasts - Toolkit"), contract no. 603458, and by the Technical University of Munich through the Global Challenges for Women in Math Science award. We also gratefully acknowledge the valuable feedback of the anonymous reviewers.

\FloatBarrier
\bibliography{CE18}

\end{document}